\documentclass[sigconf]{acmart}
\usepackage[utf8]{inputenc}
\usepackage{xspace}
\usepackage{balance}
\usepackage{graphicx}
\usepackage{float}
\usepackage{lipsum,multicol}
\usepackage{amsmath}
\usepackage{enumitem}
\usepackage{bm}
\usepackage{algorithm}
\usepackage{algpseudocode}
\usepackage{subfigure}
\algrenewcommand\textproc{}
\usepackage[colorinlistoftodos,prependcaption,textsize=medium]{todonotes}
\usepackage{booktabs}
\newcommand{\hidecomment}[1]{}
\newcommand{\hide}[1]{}
\PassOptionsToPackage{hyphens}{url}\usepackage{hyperref}

\setcopyright{rightsretained}

\newcommand{\myparagraph}[1] { \noindent {\textbf {#1}} }

\newcommand{\website}{website\xspace}
\newcommand{\websites}{websites\xspace}
\newcommand{\Website}{Website\xspace}
\newcommand{\Websites}{Websites\xspace}

\newcommand{\webpage}{web page\xspace}
\newcommand{\webpages}{web pages\xspace}

\newcommand{\Webpages}{Web pages\xspace}

{\begin{list}{$\bullet$}{%
        \setlength{\labelsep}{2pt}\setlength{\leftmargin}{10pt}%
        \setlength{\labelwidth}{0pt}%
        \setlength{\listparindent}{0pt}}}
{\end{list}}
\renewcommand{\paragraph}[1]{\vspace{0.1cm}\noindent \textbf{#1}}
\newcommand{\paragraphbf}[1]{\noindent \textbf{#1}}

\copyrightyear{2019}
\acmYear{2019} 
\setcopyright{iw3c2w3}
\acmConference[WWW '19]{Proceedings of the 2019 World Wide Web Conference}{May 13--17, 2019}{San Francisco, CA, USA}
\acmBooktitle{Proceedings of the 2019 World Wide Web Conference (WWW'19), May 13--17, 2019, San Francisco, CA, USA}
\acmPrice{}
\acmDOI{10.1145/3308558.3313709}
\acmISBN{978-1-4503-6674-8/19/05}

\usepackage{balance}

\begin{document}
\title{Bootstrapping Domain-Specific Content Discovery on the Web}

\author{Kien Pham}
\orcid{}
\affiliation{%
  \institution{New York University}
}
\email{kien.pham@nyu.edu}

\author{Aécio Santos}
\orcid{https://orcid.org/0000-0002-5124-7770}
\affiliation{%
  \institution{New York University}
}
\email{aecio.santos@nyu.edu}

\author{Juliana Freire}
\orcid{https://orcid.org/0000-0003-3915-7075}
\affiliation{%
  \institution{New York University}
}
\email{juliana.freire@nyu.edu}

\begin{abstract}
  The ability to continuously discover domain-specific content from the Web is critical for many applications.
  While focused crawling strategies have been shown to be effective for
  discovery, configuring a focused crawler is difficult and time-consuming.
  Given a domain of interest $D$, subject-matter experts (SMEs) must
  search for relevant \websites and collect a set of representative
  \Webpages to serve as training examples for creating a classifier that
  recognizes pages in $D$, as well as a set of pages to seed the crawl.
  In this paper, we propose DISCO, an approach designed to bootstrap
  domain-specific search. Given a small set of \websites,
  DISCO aims to discover a large collection of relevant \websites.
  DISCO uses a ranking-based framework that mimics the way users search
  for information on the Web:
  it iteratively discovers new pages, distills, and ranks them. It also
  applies multiple discovery strategies, including keyword-based and
  related queries issued to search engines, backward and forward
  crawling.  By systematically combining these strategies, DISCO is able
  to attain high harvest rates and coverage for a variety of domains.
  We perform extensive experiments in four social-good domains, using data
  gathered by SMEs in the respective domains, and show that our approach
  is effective and outperforms state-of-the-art methods.
  \end{abstract}



\maketitle

\section{Introduction}
\label{sec:intro}
The ability to continuously discover content relevant to an information domain has many
applications, from helping in the understanding of humanitarian crises to countering
human and arms trafficking.
It is estimated that tens to hundreds of thousands of escort ads are
posted every day and about 75\% of reported human-trafficking survivors were
advertised online at some point~\cite{thorn-report}.
Recently,
the US law enforcement has shut down the sex marketplace
\website \url{backpage.com} due to its role in sex trafficking.
Criminals also take advantage of online markets to avoid
background checks and trade illegal weapons. According to an FBI
report, hundreds of thousands of guns slip through the federal
background-check system every year~\cite{fbi-report}. At the same
time, many sites advertise weapons for sale -- in \url{armslist.com} alone one
can find hundreds of thousands of weapon-related ads.

By collecting and analyzing domain-specific content from the Web, we
can better understand these illegal activities, generate leads, and
obtain evidence for investigations.
But doing so is challenging. Currently, subject matter experts (SMEs)
(e.g., from non-governmental organizations and government agencies)
must collaborate with computing experts and data scientists to
construct specific solutions to their problems, which include (1) creating
machine-learning models to recognize relevant content, and (2) configuring
crawlers that retrieve the information at scale.
Once these are available, it is possible to continuously 
discover relevant content by using re-crawling  strategies~\cite{pham@wsdm2018,dasgupta@www2007,santos@tempweb2016}.
However, both of these tasks require users to obtain a sufficiently large
number of pages
and finding these pages is difficult.

In this paper, we address this key problem in bootstrapping
domain-specific search: how to effectively discover relevant
\websites given a small seed set provided by SMEs. 
A natural approach for discovering relevant \websites is to use
commercial search engines, such as Google and Bing, and issue
keyword queries or  search for similar sites using ``related'' search.
This requires SMEs to go through an iterative process in which they
search, distill the results, and use the information gathered to
formulate additional queries. This process is cumbersome and inefficient for
large-scale information gathering tasks.  
Different approaches have been proposed to tackle this problem.
The Domain Discovery Tool (DDT)~\cite{krishnamurthy@kdd2016} aims to
simplify the process of constructing classifiers for a given
domain. To help users discover relevant pages and train a domain classifier,
it provides an easy-to-use interface that summarizes search results
and helps users formulate new search queries.
While
the tool is effective, it was designed with a user-in-the-loop
philosophy and thus requires substantial user effort to perform the
discovery operations and review the results.
Methods such as forward and backward
crawling~\cite{barbosa@ijcnlp2011} and DEXTER \cite{disheng@vldb2015}
have been be proposed to automate discovery. Because they
rely on classifiers,  they need to be adapted
for different domains. Besides, these methods fall short of achieving high
precision and recall in domains where relevant \websites are not
highly connected.  
To the best of our knowledge, there has been no systematic comparison
of these methods for different domains, therefore whether they are
effective for website discovery is still an open question.

Several focused crawling and discovery techniques could potentially be
adapted for this problem.  However, they all rely on the availability
of an accurate domain-specific classifier~\cite{ester@vldb2004,
  vieira@wwwj2016, barbosa@ijcnlp2011}. 
This is an unrealistic assumption for the many application scenarios
where experts must start with a small set of relevant \websites, since
a small sample is unlikely to be sufficient construct an accurate
classifier.
These techniques use the classifier to select discovered pages that
are relevant, and use content of these pages to train other models
that directly control future search, e.g., ordering
unvisited URLs and generating search queries. 
Therefore, a weak classifier would gradually degrade the effectiveness of
discovery.

To address this challenge, we propose DISCO, a new method for \website
discovery  that does not depend on an accurate domain-specific
classifier.
Given a small number of sample pages, DISCO automatically discovers
additional pages that can be used both to construct domain classifiers
and to serve as seeds for focused crawlers.
The strategy adopted by DISCO was inspired by how users follow the
search process: it searches for pages on the Web and distills the
results in an iterative fashion.
Different operations can be used to discover new pages, including
keyword-based and related queries issued to search engines, and backward
and forward crawling.  
However, the effectiveness of an operation varies across  domains.
For example, in our experiments (Section~\ref{sec:expr}), we observed
that backlink search performs the best for human trafficking
domain, but is much less effective for the others. 
Based on this observation, unlike previous works~\cite{barbosa@ijcnlp2011,vieira@wwwj2016},
DISCO applies multiple search operations.
To systematically select the search operation as the discovery
progresses, we propose an algorithm that uses the multi-armed bandit
strategy~\cite{auer_ucb1}.  This makes DISCO adaptive both within and
across domains, and leads to higher harvest rates and coverage
compared to a fixed strategy.

To distill the results, in the absence of a model that recognizes
relevant pages, DISCO approximates how users select relevant pages by
ranking them. Given a set of discovered pages, it ranks higher pages
that are more similar to the input sample set.  The challenge here
lies in designing a strategy that is effective. We carried out an
empirical study of several ranking approaches and found that none of 
the approaches is uniformly better for all domains. We propose a new
ranking function that, by combining the results of multiple functions, 
produces a robust ranking.

\myparagraph{Contributions.}
Our contributions can be summarized as follows:
\begin{itemize}[noitemsep,nolistsep,leftmargin=5mm]
\item We propose DISCO, a framework that automatically discovers
  relevant \websites requiring only a small set of example \websites.
  To the best of our knowledge, ours is the first approach that does
  not require an accurate classifier to bootstrap domain discovery.
\item To distill results, we propose an ensemble ranking approach that combines multiple
  independent ranking functions, and show that it consistently attains higher precision
  than the independent functions.
\item We incorporate different search operations in our framework and
  propose an algorithm that uses the multi-armed bandits strategy to
  select the best search operator at each iteration.  We also perform
  a systematic evaluation of the performance of the different
  operations in multiple domains and show that, by using the
  multi-armed bandits-based algorithm, our method is able to attain
  high harvest rates.
\item We perform extensive experiments in multiple social-good domains
  using real data provided by SMEs from the corresponding
  domains.  We report 
  experimental results which show that DISCO obtains 300\% higher harvest rate
  and coverage compared to state-of-the-art techniques for domain discovery.
\end{itemize}

%
%

\paragraph{Outline.} The remainder of this paper is structured as
follows.  We discuss related work in Section~\ref{sec:related}.
In Section~\ref{sec:overview}, we define the \website discovery and
ranking problems and give an overview of the DISCO framework.  In
Section~\ref{sec:rank_disco} we describe the ranking and discovery
components of the framework.  We present the results of our
experimental evaluation in Section~\ref{sec:expr}, and conclude in
Section~\ref{sec:conclusion}, where we outline directions for future
work.

\section{Related Work}
\label{sec:related}

Several techniques have been proposed to discover domain-specific web
content. These can be categorized in the following two groups:
\textit{Search-based Discovery} techniques~\cite{disheng@vldb2015,
  vieira@wwwj2016, wang@WWW2018} which rely on search engine APIs (e.g.,
Google, Bing, and Alexa APIs) to find \webpages similar to given
keywords or \webpages; and
\textit{Crawling-based Discovery} techniques~\cite{murata@WWW,
  barbosa@ijcnlp2011, barbosa@ICWE}, which use the Web link structure
to explore new content by automatically downloading and recursively
following links extracted from the discovered \webpages.
Note that while some of these techniques were proposed to discover
specific \webpages, they can be extended to discover \websites as
well.
To the best of our knowledge, as we discuss below, none of the prior
work supports all different discovery techniques, and most require an
accurate domain-specific classifier.

\subsection{Search-Based Discovery}
There are two main search-based discovery operators supported by
search engine APIs: \textit{keyword
  search} and \textit{related search}.  Most of the existing discovery
techniques use keyword search since related search is only supported
by two search engines,  Google and Alexa.

Earlier work~\cite{dean@SD1999, murata@WWW} used both content and Web
link structure to compute the relatedness between \webpages.
Vieira et. al.~\cite{vieira@wwwj2016} proposed a system that uses
relevance feedback to gather seeds to bootstrap focused crawlers. It
submits keyword search queries to Bing; extracts keywords from the result pages
classified as relevant for the focus domain; and uses these keywords
to construct new search queries.
Their system shares some of our goals and
their findings support our design decisions. They show that submitting multiple
short queries and retrieving a small list of results leads to more
relevant pages than submitting long queries and extracting
a large number of results.
\hide{
 Also, the new queries are generated using the
documents retrieved from the previous queries.  Second, it uses a
domain-specific classifier to continuously refine queries through a
pseudo-relevance feedback approach.
}
The major drawback of this approach is that, when the classifier is
not accurate, the search results quickly diverge from the target
domain.  Furthermore, the system is specifically designed to use with
keyword search, therefore, does not support other search techniques.

Disheng et. al.~\cite{disheng@vldb2015} presented a discovery pipeline
that uses keyword search to find websites containing product
specifications (e.g., cameras and computers).  While they also utilize
backward search, they showed that the relevant \websites are mainly
discovered by searching with product identifications extracted from
previous discovery iterations.  However, using product identifications
as search keyword is specific to product search domain and therefore
not applicable to other domains.

Wang et. al.~\cite{wang@WWW2018} focused on the claim-relevance
discovery problem that, if solved, can help to identify online
misinformation.  They proposed a system that extracts salient keywords
from fact-checking articles and uses them to search for candidate
documents.  Unlike in our search strategy, their search phase is not
iterative, since they do not aim to attain high coverage of relevant documents.

\subsection{Crawling-Based Discovery}
Forward and backward crawling are two types of crawling-based
techniques explored in the literature.
In forward-crawling, links are extracted from the discovered
\webpages.  Several forward-crawling approaches  been proposed
to discover domain-specific \webpages and \websites~\cite{soumen1999,
  soumen2002, barbosa@www2007, ester@vldb2004}.  The most prominent
approach in this group are the \textit{focused crawling} methods that
employ online learning policies ~\cite{soumen2002, barbosa@www2007,
  meusel@cikm2014}.  They use two classifiers to focus the
search: the critic, a classifier that categorizes \webpages as
relevant or irrelevant to the domain; and the apprentice, a classifier
that learns to identify the most promising links in a domain-relevant
page in an online fashion.  Note that while the apprentice classifier
is automatically learned, the critic is a classifier given as input to
the system and requires a substantial number of positive and negative
examples to be trained on.  In contrast, our approach does not require
an input classifier, just a small set of relevant pages.
Compared to search-based techniques, forward crawling is more
scalable, however, it works best for domains where relevant content is
well-connected.

In contrast to forward crawling, backward crawling (or reverse
crawling) discovers new pages through backlink search. Since the Web
is unidirectional, it must be done through search engine APIs that
expose results of previous large-scale crawls.
Early work proposed the use of backlink search to discover Web
content~\cite{diligenti2000, chakrabarti@cnj1999}.
More recently, Barbosa et. al.~\cite{barbosa@ijcnlp2011} proposed a
crawling strategy that combines both backward-crawling and
forward-crawling techniques to locate bilingual websites.  It employs
two link classifiers, trained with the content of visited pages and
their relevance to the target domain, to determine the visiting orders
of the links discovered by the two techniques.  As a result, the
efficiency of this strategy highly depends on the domain classifier.
In addition, this approach has the same limitation we discussed above,
in that the crawler becomes ineffective when the classifier is not
accurate.
In later work, Barbosa~\cite{barbosa@ICWE} proposed a similar crawling
architecture for discovering forum \websites.

Approaches that focus on real-time
discovery~\cite{rogstadius@IBMJ2013, shao@WWW2016,
  zubiaga@ACMSurvey2018} have taken advantage of social media platforms
(e.g., Twitter) to monitor and track domain-specific content such as
misinformation~\cite{shao@WWW2016}, activity related to natural
disasters~\cite{rogstadius@IBMJ2013}, and
rumors~\cite{zubiaga@ACMSurvey2018}. Although these techniques mainly
target social-media feeds, they could potentially be used to discover
\websites by extracting links present in the feeds. Discovery from
social media could be integrated in our approach as an alternative
discovery operation.

\section{Problem and Solution Overview}
\label{sec:overview}

\subsection{Problem Definition}
\label{sec:problem}

\paragraph{\Website Discovery Problem.}
Let $D$ represent the domain of interest and $w$ denote a \website
represented as one of more \webpages.
If $w$ is relevant to the domain, we say that $w \in D$.
Given a small set of \websites $S \subset D$ (seeds), our
goal is to efficiently expand $S$ by finding additional \websites $w$
that belong to $D$.

\paragraph{\Website Ranking Problem.}
In the absence of a model
that recognizes pages in $D$, we approximate the problem to make it
depend only on the set $S$ of seeds by framing it as a ranking task:
given $S$, the goal is to discover new \websites that are \emph{similar} to $S$.
More formally, given set of discovered \websites $D_+$, the problem
consists of  ranking each site $w$ in $D_+$ according to
its similarity with $S$ -- this is equivalent to the likelihood of $w$
being in $D$.
The intuition is that the closer $w$ is to $S$, the more likely it is that
$w \in D$.  

By ranking the discovered \websites based on their similarity with
$S$, we help to reduce the SMEs effort in reviewing and labeling
examples to build a domain-specific classifier and selecting important
\websites for the re-crawling process.
A ranking strategy in this scenario should ideally have two important features:
high precision in the top of the ranked list and low false negative in the tail of the ranked list.
The former feature is important for the \website discovery operation
to be effective since the top-ranked websites are fed back into the
discovery process. The latter is useful for reducing the SMEs effort
of reviewing the final ranked list output by the system: the fewer
relevant documents in  the tail of the list, the fewer documents need
to be reviewed by the SMEs.  In our experimental evaluation in
Section~\ref{sec:expr}, we use metrics aligned with these desired
features to evaluate different ranking functions.

\begin{figure}[t]
  \centering
  \includegraphics[height=0.23\textheight]{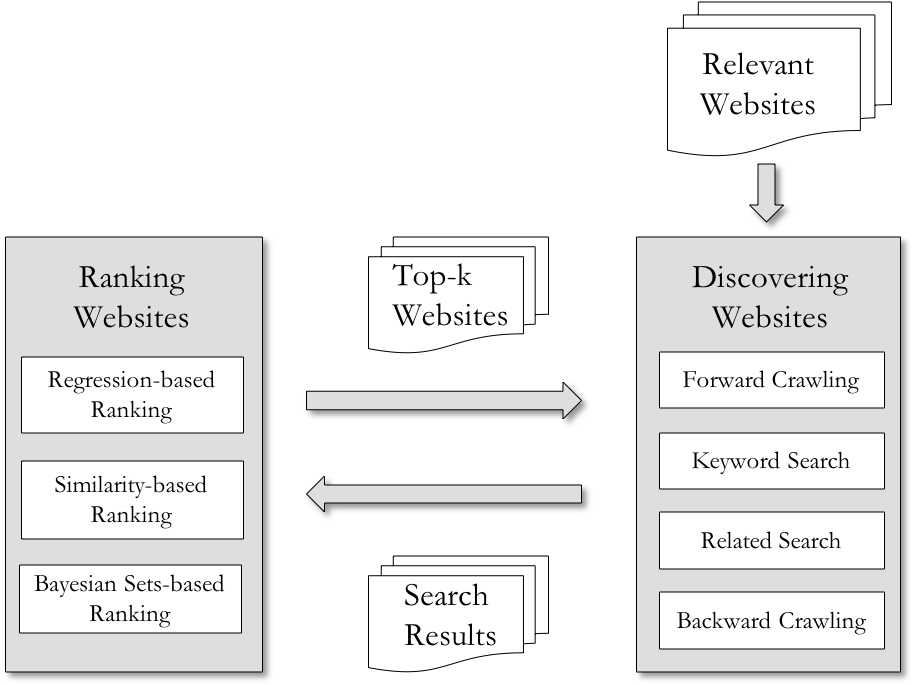}
  \caption{Architecture of DISCO.}
  \label{fig:framework}
\end{figure}

\subsection{Framework Overview}

The design of DISCO was inspired by the iterative process that SMEs usually
follow to discover domain-specific \websites.
They start by submitting queries to a search engine. Then, they review
and evaluate the search results based on their knowledge about the
domain. The new knowledge they acquire in this process is used to
refine and formulate new search queries.
In DISCO, we replace the manual evaluation with the
\website ranking component and  use search engine APIs to automate the search operations.
Figure~\ref{fig:framework} depicts the high-level architecture of the DISCO framework.
It consists of two main components: \textit{\website ranking} and \textit{\website discovery}.
To incorporate different search engine APIs and ranking techniques, we
standardize the input and output of the two components.  Specifically,
the \website discovery component takes a list of relevant
\websites as input and returns a list of discovered \websites after performing
the search operations.  Similarly, the input and output of the \website
ranking component consist of lists of discovered \websites so far and the
corresponding list of ranked \websites, respectively.  
This design
unifies multiple search operators in a single framework, and hence, it
allows us to fairly compare their efficiency across different domains
(see Section~\ref{sec:expr}).
Algorithm~\ref{alg:discovery} outlines the discovery process carried
out by DISCO.
It starts
with the seeds and iteratively searches, ranks and selects the top
results.
The
\textit{select\_discovery\_operator()} function in line 6 returns one
of the four discovery operators: forward crawling, backward crawling,
keyword search and related search.

\begin{algorithm}[t]
\caption{\Website Discovery}
\label{alg:discovery}
\begin{algorithmic}[1]
\Procedure{Discovery($seeds$)}{}
\State $results = \emptyset$
\State $ranked\_results = \emptyset$
\State $topk = seeds$
\While {stop condition}
\State $op = select\_discovery\_operator()$
\State $search\_results = op.search(topk)$
\State $results = results \cup search\_results$
\State $ranked\_results = rank(results)$
\State $topk = get\_top\_k(ranked\_results)$
\EndWhile
\State \textbf{return} $ranked\_results$
\EndProcedure
\end{algorithmic}
\end{algorithm}

\section{Ranking and Discovery}

\label{sec:rank_disco}
In this section, we present the details of the \website ranking and
discovery components of DISCO. Then, we present a technique to
efficiently select the discovery operator based on the multi-armed
bandits algorithm.

\subsection{\Website Ranking}
\label{sec:ranking}
The \website ranking problem can be cast as a conventional ranking problem in information retrieval, where
$S$ represents the query, and the goal is to rank the \websites in $D_+$ according to its similarity to the query.
The query here consists of a very long document containing all the \websites in $S$,
which makes the traditional ranking models (e.g., BM25~\cite{robertson-bm25@FTIR2009}) less effective.
Furthermore, merging all the \websites in $S$ can potentially blur the
distinct domain-specific features that exist in individual \websites.
To avoid this problem, we investigate other ranking approaches that consider \websites in $S$ individually.
We then propose a new ranking function, inspired by ensemble learning methods, 
that combines the scores of the existing approaches.

\paragraph{\Website Representation.}
The first step of any ranking function design is to represent the candidate documents, which are \websites in our context.
A \website can be represented as a single or multiple \webpages. We
consider both options and selected the latter for several reasons.
To use multiple pages to represent a site, we need to have an
effective procedure to select representative pages in the sites. This
task is challenging, especially when we do not have ground truth data.
Random selection tends to result in a large number of \webpages that
are not good representative of the site and that lead to noise that
negatively affects the ranking.
The single-\webpage representation can help circumvent
this issue if we rely on the assumption that every \website contains
at least one \webpage that can serve as an effective representative
of the \website content.
As a result, the score of a \website can depend only on the score
of its most relevant \webpage discovered.
Furthermore, the single-\webpage representation greatly simplifies the implementation 
of the framework and improves its performance. 
For this representation, we do not need a separate process to retrieve
and manage additional pages for each newly discovered \website,
and discovered \webpages can be processed independently.
This makes the framework ready for parallelization.

In fact, we experimentally compared  selection strategies using
multiple \webpages representation against the single-page
representation, and the former led to better rankings than the latter for all domains
we considered.
We considered two different approaches for the multiple-page
representation.
One 
performs a breadth-first-search crawling starting from the
home page, which leads to both relevant and irrelevant pages. The
other uses in-site search via search engine APIs (i.e., using
queries with the format ``site:keyword``). This method, however, is
keyword-sensitive and tends to retrieve false positive pages.

Several ranking function have been proposed in the literature. In what
follows, we describe the ones we considered. Details about their
performance are discussed in Section~\ref{sec:expr}.

\subsubsection{Regression-Based Ranking}
This approach learns the ranking function using regression-based
techniques and computes a score for each \website in $D_+$.  The
challenge is that only positive examples (i.e., $S$) are given, and
thus additional negative examples must be obtained or unconventional
techniques such as PU learning~\cite{bingliu@ICDM} or novelty detection~\cite{scholkopf@nc2001} must be applied.  We investigate three regression techniques in
this direction and describe how they can be applied in our scenario.

\paragraph{Binomial Regression.}
In order to use binomial regression (i.e., logistic regression), we
need both positive and negative examples.  Therefore, we obtain
negative examples by randomly selecting \webpages from a public Web
corpus.  Since the Web corpus is relatively large compared to the
number of samples pages, there is only a very small chance of
selecting a \webpage relevant to the domain.  Given $S$ and the
unlabeled \webpages as negative examples, we train a binomial
regressor and use it to compute ranking scores.

\paragraph{Positive and Unlabeled Example Learning (PU Learning).}
A well-known alternative to binomial regression method, given positive
and unlabeled examples, is PU learning~\cite{bingliu@ICDM}.
\citet{bingliu@ICDM} proposed biased SVM, a PU learning technique that frames the PU
Learning problem as a constrained optimization problem.
They experimentally showed that biased SVM outperforms other
2-steps PU learning approaches.  Therefore, we choose this technique
to learn the ranking function.

\paragraph{Novelty Detection.}
Another line of research that does not require negative examples is
novelty detection~\cite{scholkopf@nc2001, markou@2003}.  
Scholkopf et. al.~\cite{scholkopf@nc2001} extend SVM (support vector
machine) to learn a function that is positive on the positive examples
and negative on its complement.  Intuitively, it computes the minimal
region in feature space that encloses the positive examples.

\subsubsection{Similarity-Based Ranking}
Similarity-based ranking, inspired by the $k$-nearest neighbors
algorithm, computes the ranking score of a \website according to the average of its
similarities to all \websites in $S$. Let $Sim(w_i, w_j)$ be
a similarity function between the \websites $w_i$ and $w_j$, the score of
a \website $w \in D_+$ is computed as follows:
\begin{equation}
Score(w) = \frac{1}{|S|}\sum_{w_s \in S}Sim(w, w_s)
\end{equation}
We consider Jaccard and Cosine as similarity functions.
These are two well-known similarity measures for textual content
that use distinct input representations, which might encapsulate
different ranking advantages. We use the vector space
model and binary vector to represent \websites for computing Cosine
and Jaccard similarity respectively.  Since these similarity measures
use different representations of the \websites, they might expose
different characteristics in the ranking task.  Let $x, y$ be the
representation of \websites $w_x, w_y$, the Jaccard index and Cosine
similarity between them are computed as follows:
\begin{equation}
Jaccard(x, y) = \frac{x \cap y}{x \cup y}
\end{equation}
\begin{equation}
Cosine(x, y) = \frac{\sum_{i=1}^{n}x_iy_i}{\sqrt{\sum_{i=1}^{n}x_i^2}\sqrt{\sum_{i=1}^{n}y_i^2}}
\end{equation}
\subsubsection{Bayesian-Sets-Based Ranking}
Bayesian Sets (BS) is a Bayesian inference technique introduced by
Ghahramani and Heller in~\cite{zoubin@NIPS} that performs set
expansion -- it generates high-precision sets of related items.
Specifically, given a set of related items that belong to a class $C$,
BS scores a new item by estimating its normalized marginal probability that the item belongs to $C$.
By this definition, BS can be directly applied to our ranking problem by considering the related items as seeds $S$ and class $C$ as domain $D$.
Accordingly, the score of a \website $w$ is computed as the probability of $w$ given that $S$ is observed:

\begin{equation}
Score(w) = \frac{P(w|S)}{P(w)}
\end{equation}

\subsubsection{Ensemble Ranking Function}
While many alternatives are possible for ranking functions, none of
them is uniformly better for all domains, as we experimentally show in Section~\ref{sec:expr}.
Thus, ideally, we should be able to combine them.
Ensemble learning is a paradigm that combines multiple
machine-learning models and has been shown to be effective
at reducing their variance.
In our context, each ranking function captures a different set of features, hence their combination 
has the potential to improve the robustness and reduce the variance of the ranking.
Inspired by this, we propose a new ranking function that combines the results from multiple ranking functions.
Specifically, let $F = \{f_i\}$ be the list of the ranking functions and
$f_i(w)$ denote the position of $w$ in the ranked list that $w$ belongs to.
The new ensemble score of $w$ is computed as the mean of these position values:
\begin{equation}
Score(w) = \frac{\sum_{f_i \in F}f_i(w)}{|F|}
\end{equation}
Another option could be to directly consider the similarity scores as $f_i(w)$.
However, these scores must be calibrated for the ensemble function to be effective.
Our similarity functions do not produce naturally calibrated scores
(i.e., probabilities), hence, we decided to use rank positions for
combining the results.
We also considered using other rank combination methods that have been
proposed in previous work, e.g., reciprocal rank fusion~\cite{cormack@SIGIR2009}.
However, we found that the mean function performed comparably well
in our case and at same time it was faster.

\subsection{\Website Discovery Operators}

The goal of the discovery operators in our context is to discover new 
\websites,
therefore we make use of existing discovery methods in a way that maximizes this objective.
We investigate four different discovery techniques:
forward crawling, backward crawling, keyword search, and related search.
To incorporate these discovery methods in our framework, we designed them in such a way that they take the same input, the top-$k$ most relevant \websites,
and produce the same output, a list of new \websites.
In what follows, we describe in detail each of these methods.

\paragraph{Forward Crawling.}
The forward crawling operator discovers new \websites through outlinks
extracted from the HTML content of the input ~\webpages.  Since the
objective is to maximize the number of relevant \websites, the
operator only returns the outlinks for \websites have not been
discovered in the previous iterations.

\paragraph{Backward Crawling.}
The backward crawling operator discovers new \websites by performing two operations: backlink search and forward crawling.
In backlink search, it calls search engine APIs to obtain backlinks
for  the input \webpages.
Since the input is relevant to the domain, the discovered backlinks
are likely to serve as hubs pointing relevant pages. This phenomenon
has been referred to as co-citation in the literature~\cite{kumar@1999}.
Then, we can discover additional relevant \websites by extracting
outlinks from the hubs and doing forward crawling.

\paragraph{Keyword Search.}
The keyword search makes use of search engine APIs to search for
\websites relevant to the keywords.  The problem is how to obtain
domain-specific keywords from the input \websites that lead to more
relevant \websites.  The body text extracted from the input pages 
contains good keywords that are representative of the domain, however,
they often co-occur with terms that may not be representative.
Our observation is that the keywords extracted from the metadata tags
(i.e., description and keyword) are significantly less noisy than the
ones extracted from the body text.  For this reason, they are
particularly effective for searching for relevant \websites.
For each page, we extract and tokenize the metadata from the
description and keywords tags.  We select the most frequent tokens as
search keyword candidates.
The problem is that, even though many keywords are relevant to the
domain, do not comprise sufficient information to retrieve relevant
\websites.  For example, the keywords \textit{pistol}, \textit{texas},
and \textit{sale} are related to the \textit{weapon forums} domain,
but more contextual information is needed to build search queries that
find additional relevant \websites.
To address this challenge, we combine these keywords with terms that
are good descriptors of the domain.
For example, in the \textit{weapon forums} domain, given
\textit{gun forum} as seed keyword, the corresponding search terms of
the previous keywords are \textit{gun forum pistol}, \textit{gun forum
  texas} and \textit{gun forum sale}.  Note that we only need one seed
keyword for each domain, so this should not add any burden to the
SMEs.

\paragraph{Related Search.}
The related search operator takes a \website URL as input and returns a list of related \websites.
To our knowledge, only Google Search APIs and Alexa APIs support this operation.
Depending on the nature of the data that these search platforms possess and their underlying algorithms,
the results can be vastly different.
Nevertheless, it appears to be a natural fit for our objective of discovering relevant \websites.
Note also that different search providers can be easily included
as different search operators.

\vspace{-.3cm}
\subsection{Discovery Operator Selection}
\label{subsec:operator_election}
In the previous section, we have described four different operators for discovering relevant \websites,
the challenge now is how to select an operator at each iteration in Algorithm~\ref{alg:discovery}
that maximizes the rate of discovering new relevant \websites.

This is a non-trivial task for several reasons.
First, it is not clear which search operator performs the best across multiple domains.
In our experiments, described in Section~\ref{sec:expr}, backward crawling performs better
than other operators for website discovery in the Human Trafficking domain,
but it performs poorly in other domains.
Furthermore, favoring only search operators that are known to be good can
leave the discovery process stuck in a connected region of the Web.
On the other hand, using different discovery operators can diversify the discovery process,
and eventually improve the rate of harvesting relevant \websites.

To address this problem, we propose a strategy to balance the trade-offs between different discovery operators
using multi-armed bandits (MAB), particularly UCB1 algorithm~\cite{auer_ucb1}.
MAB is known as an effective means to solve the exploitation and exploration problem.
The goal of this strategy is to select bandit arms with the highest accumulated reward and 
to penalize the ones that were overly operated in the past.
To accomplish this, we need to define two variables in the UCB1 algorithm:
the bandit arm ($op$) and the reward of its operation ($\mu_{op}$).
We model each discovery operator as a bandit arm.
Assume that each time $t$, when an arm $op$ is selected, it discovers $n_{op,t}$ \websites.
According to the UCB1 algorithm, the score of $op$ at time $t$ is defined as:

\vspace{-.3cm}
\begin{equation} \label{ucb_score}
Score_{op,t} = \overline{\mu_{op}} + \sqrt{\frac{2ln(n)}{n_{op,t}}}
\end{equation}
In this equation, $n$ is the total number of \websites retrieved by all operators.
At each iteration, the algorithm selects the operator with highest score to perform the discovery.

The remaining challenge is to define the reward of an operator so that
it is proportional to the number of relevant \websites it discovered.
Although we do not know the relevance of the discovered \websites,
we can approximate them by their ranking scores computed by the \website ranking component (Section~\ref{sec:ranking}).
Let $pos_i$ denote the position of a \website $i$ in the ranked list.
The normalized reward of $i$ is $1-\frac{pos_i}{len}$, where $len$ is length of the ranked list.
By this definition, if $i$ is ranked at the end or top of the list, its reward is 0 or 1 respectively.
Since the goal is to maximize the number of new relevant \websites, we assign $0$ reward to a \website
that was already discovered in previous iterations, regardless of its position.
As a result, the reward of $i$ is redefined as $(1-\frac{pos_i}{len})*I_i$, where $I_i$ takes binary values:
\begin{equation}
I_i =
\left\{\begin{matrix}
0 & \text{\website $i$ was already discovered} \\
1 & \text{otherwise}
\end{matrix}\right.
\end{equation}
If a discovery operator $op$ returns $k$ \websites, its reward $\mu_{op}$ is defined as follows.
\begin{equation}
\mu_{op} = \frac{\sum_{i=1}^{k}(1-\frac{pos_i}{len})*I_i}{k}
\end{equation}
We use this reward to compute the score of an operator at time $t$ ($Score_{op, t}$) 
in the equation~\ref{ucb_score}. After each round of search, we recompute this score
and select the operator with highest score for the next round.

\section{Experimental Evaluation}
\label{sec:expr}
In this section, we evaluate the effectiveness of our discovery framework and
ranking functions through extensive experiments using real-world datasets.
We use the following domains for the experiments:
\begin{itemize}[noitemsep,nolistsep,leftmargin=5mm]
\item Human Trafficking (HT): classified ads and other escort-related \websites which 
often contain adult ads. Prominent example \websites in this domain are \texttt{backpage.com},
\texttt{craigslist.com}, and \texttt{adultsearch.com}.
\item Weapon Forums (Forum): discussion forums about firearms and weapons.
\item Weapon Marketplaces (Market): marketplace or classified ads \websites for firearms and weapons.
\item Stock Promotions (SEC): \websites that post stock promotion alerts. This domain contains information that can support investigations of micro-cap fraud.
\end{itemize}
We selected these domains for the following reasons.
First, since they are different in nature, they allow us to evaluate the robustness of our proposed 
techniques and baselines.
Most importantly, we had access to the seed \websites and the training
examples manually labeled by SMEs in these domains.  The SMEs were
analysts from government agencies and worked with us in collecting and
reviewing relevant content.
The number of seed \websites in HT, Forum, Market and SEC domains are
54, 103, 23, and 18 respectively.
These are representative \websites that contain a significant number of pages relevant
to the domains.
These seed \websites are used for the evaluation of the ranking methods in \ref{sec:rank_eval} and for the
input of the DISCO framework in Section~\ref{sec:disc_eval}.
As training examples (described in Table~\ref{table:training_stat}),
we used a 
bigger set of websites, labeled as relevant and non-relevant by SMEs
in HT, Forum and Market domains. 
We use these examples to create ground-truth classifiers for the
evaluation of the DISCO framework, which we present in Section~\ref{sec:disc_eval}.

It is worthy of note that although we were able to obtain ground-truth
to train these classifiers, labeling the pages required a very time-consuming 
process. In addition, the labeling process was not a trivial task for
the SMEs, who are familiar with the domain but not with the
intricacies of learning. For example, they did not know the
impact of labeling some confusing cases would have in the final model.
An important motivation for this work is to devise an easier-to-use and less
time-consuming process to solve this problem.

\subsection{Evaluation Metrics}
\hide{
In what follows, we describe the metrics we use to evaluate the \website discovery and the ranking problem described in Section~\ref{sec:problem}.
}
\subsubsection{\Website Ranking Metrics}
Section~\ref{sec:problem} presented two desired characteristics of good ranking
functions: high precision in the top of the ranked list and low false negative in its tail.
In order to measure these qualities, we use the following metrics:

\noindent \textbf{P@k}: computes the percentage of the relevant \websites in the top-$k$ \websites of the ranked list.
This metric captures the quality at the top of the list.

\noindent \textbf{Mean Rank}: computes mean of the ranking positions of the relevant \websites in the ranked list.
\begin{equation}
MR = \frac{1}{n}\sum_{w_i \in D_+^*}position(w_i)
\end{equation}
We do not normalize the rank by the length of the ranked list because we want to keep the absolute values for easy interpretation.
Also, in our experiment, the length of the ranked lists in the experimental domains are similar,
therefore absolute values are still valid for cross-domain comparison.

\noindent \textbf{Median Rank}: position of the relevant \website whose position after ranking is the midpoint of all the relevant \websites.
\subsubsection{\Website Discovery Metrics}
We measure the discovery efficiency using \textit{Harvest Rate} and \textit{Coverage} metrics.
In the following metric definitions, let $D_+^*$ be the set of discovered \websites that are in the domain $D$.

\noindent \textbf{Harvest Rate:} the ratio between the size of $D_+^*$ and the size of $D_+$. 
The Harvest Rate is a standard metric that has been widely used in literature to measure the efficiency of focused crawlers.

\noindent \textbf{Coverage:} we define coverage as the percentage of \websites in $D$ that are discovered.

\subsection{\Website Ranking Evaluation}
\label{sec:rank_eval}
For each domain, we randomly select 10,000 \websites from DMOZ directory\footnote{http://dmoz-odp.org/} as negative candidates.
Since the domains are rare and the size of DMOZ is large (about 4M \websites),
there is only a small chance that the selected examples are relevant to the domain.
Therefore, we can reliably consider these as negative candidates.
For each domain, we split the seed \websites into train and test sets randomly.
We merge the test set into the set of negative examples, creating a new set named the \textit{candidate set}.
Then, we rank the candidate set using the  ranking functions described in Section~\ref{sec:ranking}.
We compute the evaluation metrics based on the positions of the test set in the ranked list.

\paragraphbf{JACCARD, COSINE}: These ranking functions compute the similarity between two \websites
using JACCARD and COSINE similarity index, respectively.

\paragraphbf{BS}: This ranking function uses Bayesian sets technique.
We represent each \website as a term-frequency vector as opposed to the binary vector used in~\cite{zoubin@NIPS}.

\paragraphbf{ONECLASS}: This ranking function uses novelty detection (i.e., one-class SVM~\cite{scholkopf@nc2001}).

\paragraphbf{BINOMIAL}: This ranking function uses binomial regression.
We use logistic regression because it is fast and known for performing well on text data.
To obtain the negative examples, we randomly select \websites from DMOZ.
The number of selected examples was equal to the number of positives examples.
For the implementation of both \textit{ONECLASS} and \textit{BINOMIAL}, we used scikit-learn~\cite{scikit-learn}.

\paragraphbf{PUL}: This ranking function uses biased SVM~\cite{bingliu@ICDM}.
We use SVM light\footnote{http://svmlight.joachims.org/} as an implementation of biased SVM.
We also use DMOZ to obtain the unlabeled examples for learning this function.

\paragraphbf{ENSEMBLE}: This is our ranking function that combines all the above ranking functions, except \textit{PUL}, 
whose performance showed a high variability across different domains.

\begin{figure*}[pht]
    
  \begin{subfigure}
    \centering
    \subfigure[Market Domain]{
    \includegraphics[height=0.17\textheight]{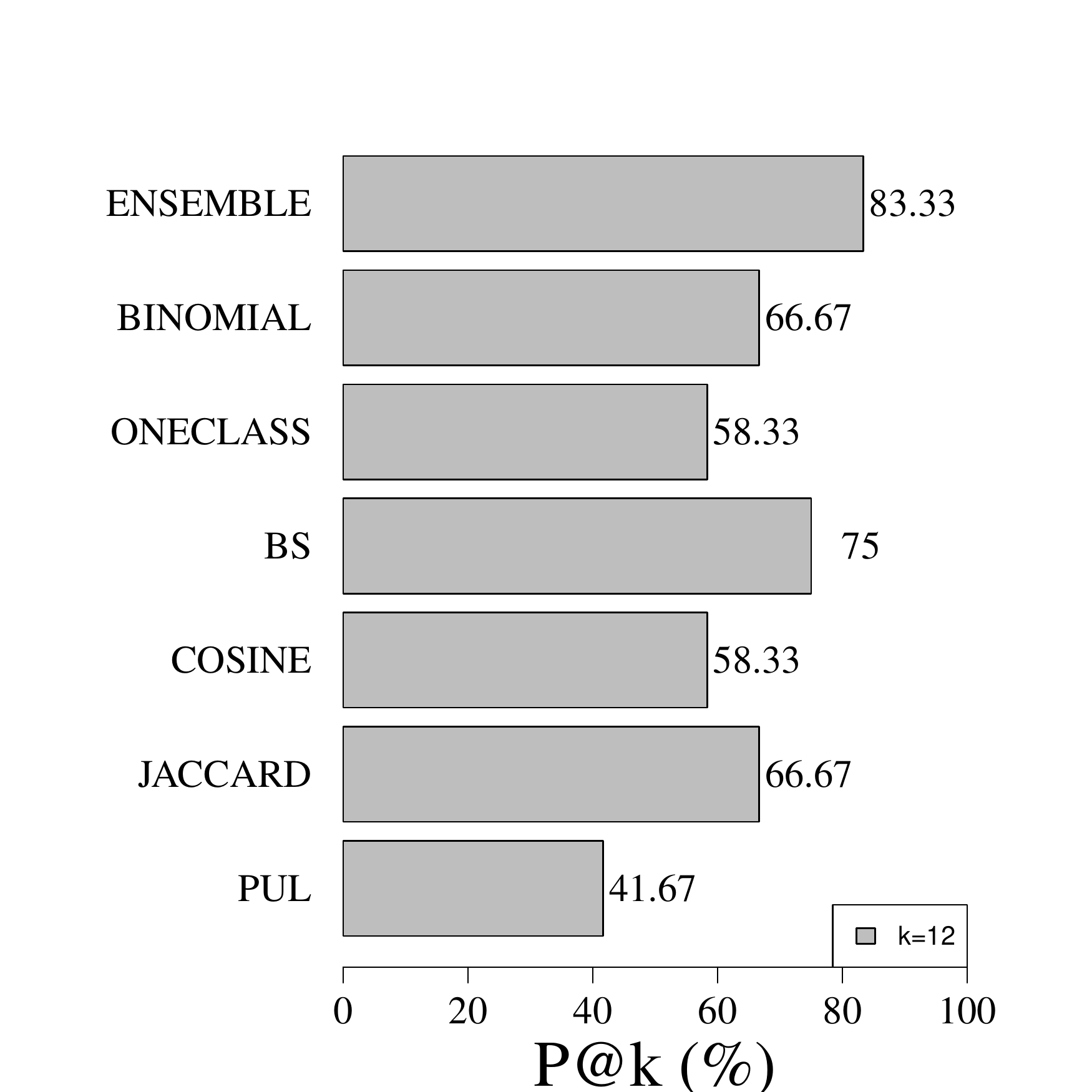}}
    \subfigure[Forum Domain]{
    \includegraphics[height=0.17\textheight]{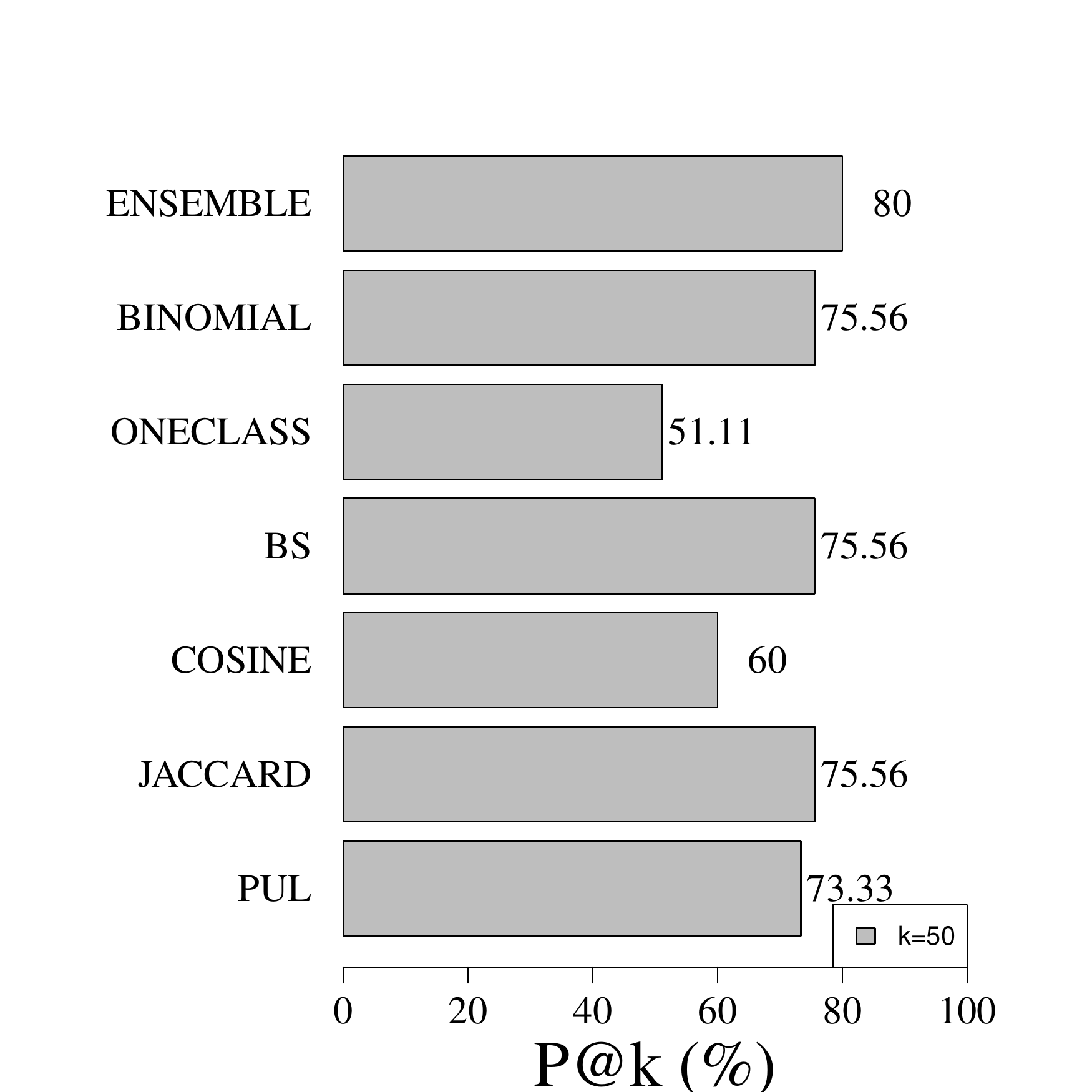}}
    \subfigure[HT Domain]{
    \includegraphics[height=0.17\textheight]{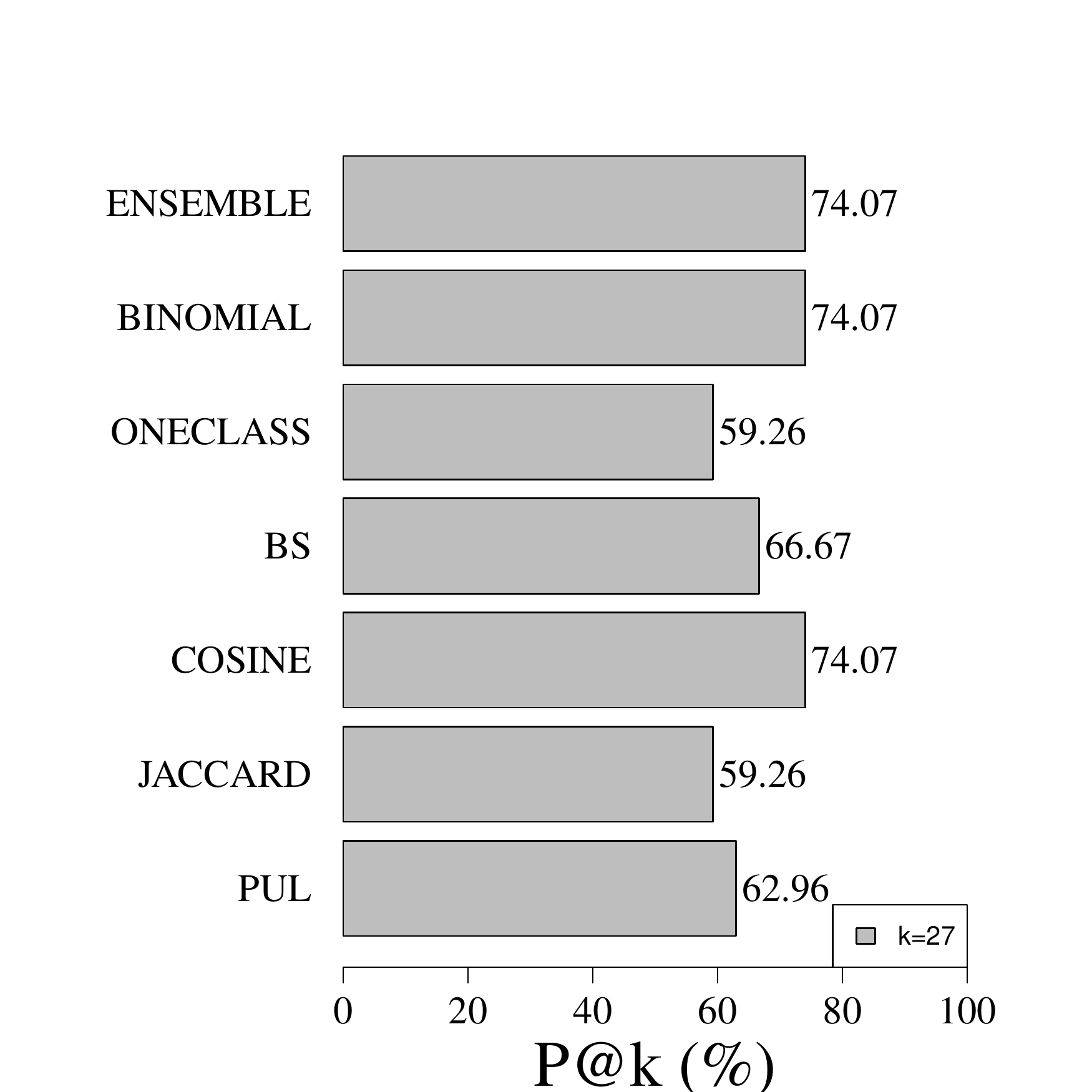}}
    \subfigure[SEC Domain]{
    \includegraphics[height=0.17\textheight]{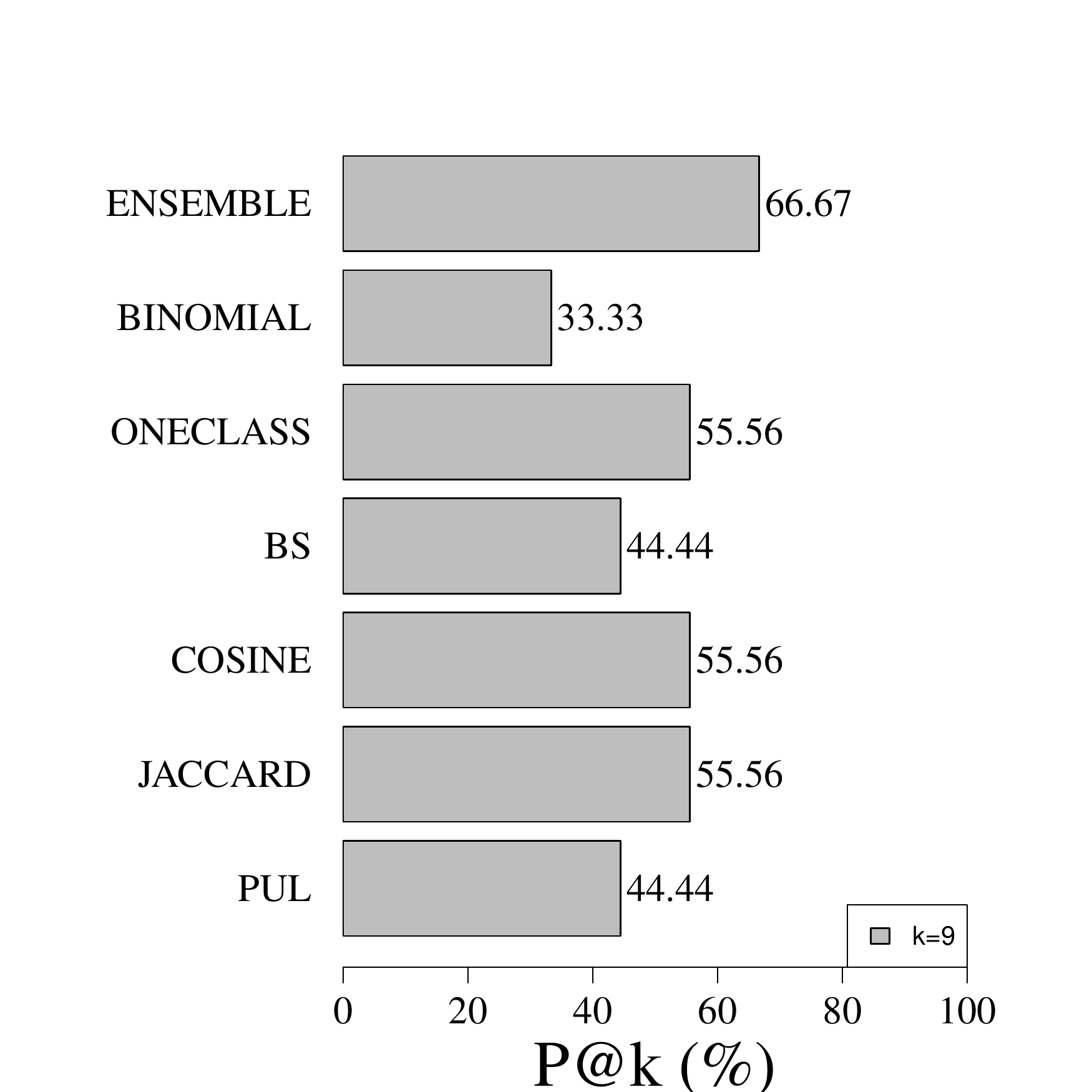}}
    \caption{Comparison of different ranking functions using P@k metric in all domains. Higher values are better.}

    \label{fig:prec}
  \end{subfigure}

  \begin{subfigure}
    \centering
    \subfigure[Market Domain]{
    \includegraphics[height=0.17\textheight]{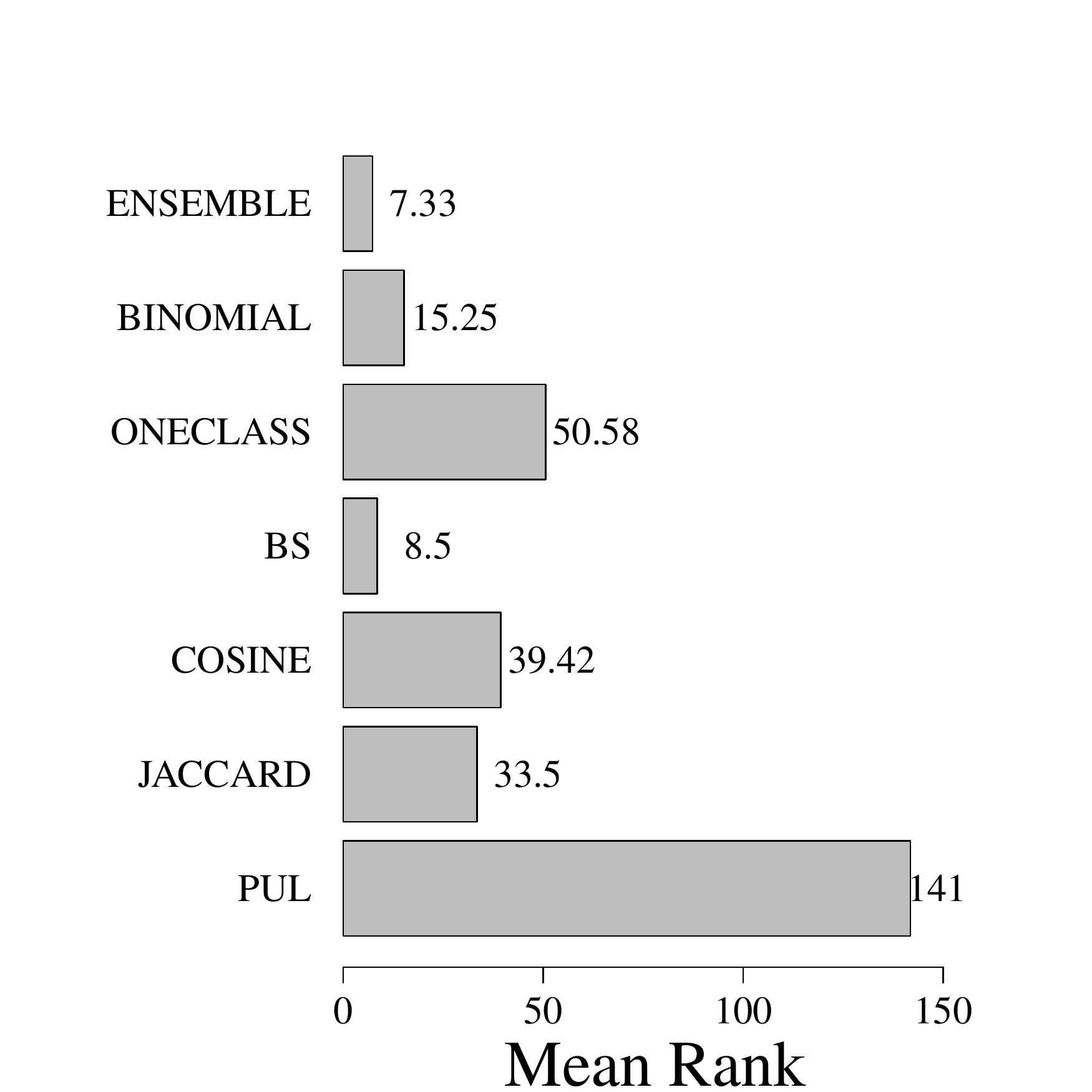}}
    \subfigure[Forum Domain]{
    \includegraphics[height=0.17\textheight]{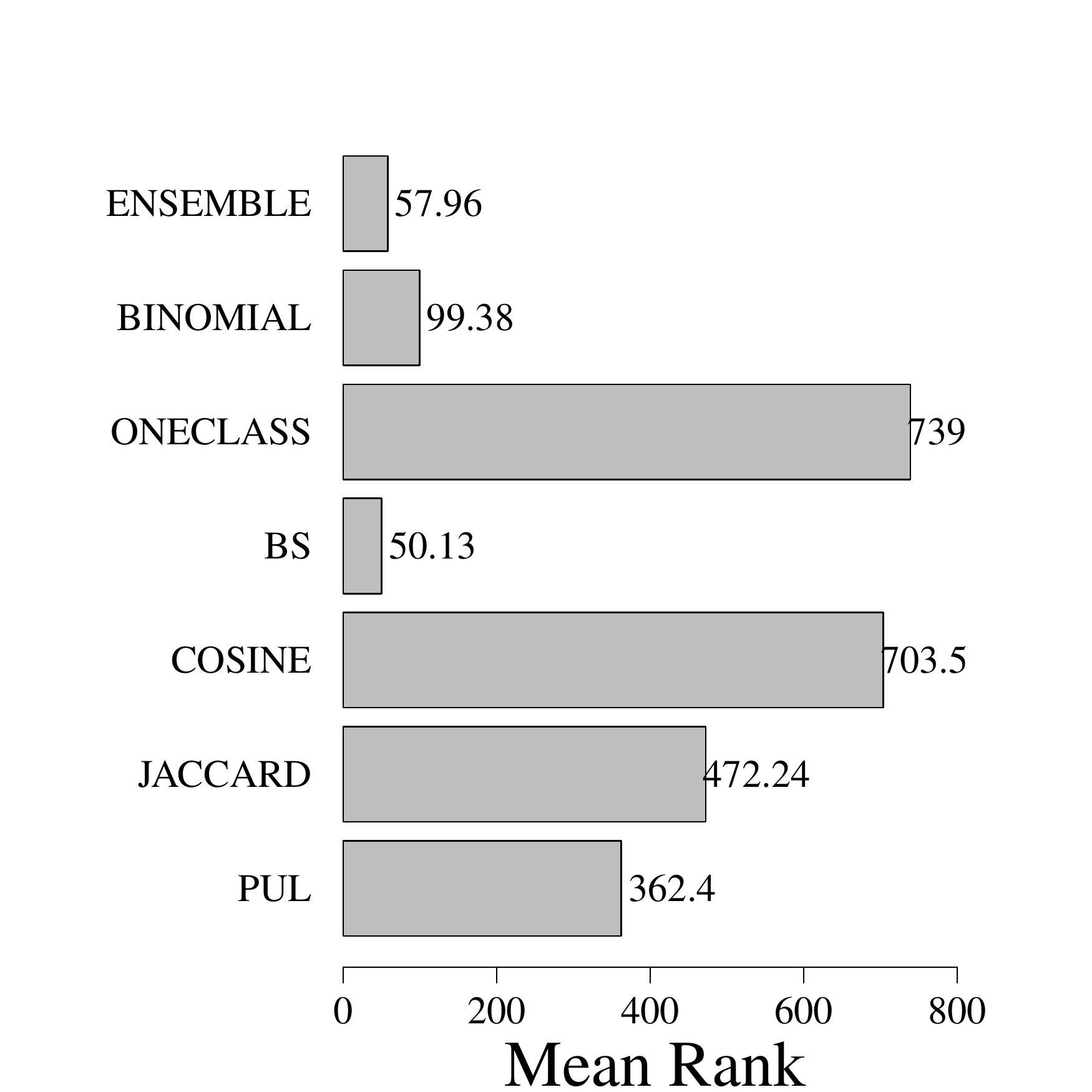}}
    \subfigure[HT Domain]{
    \includegraphics[height=0.17\textheight]{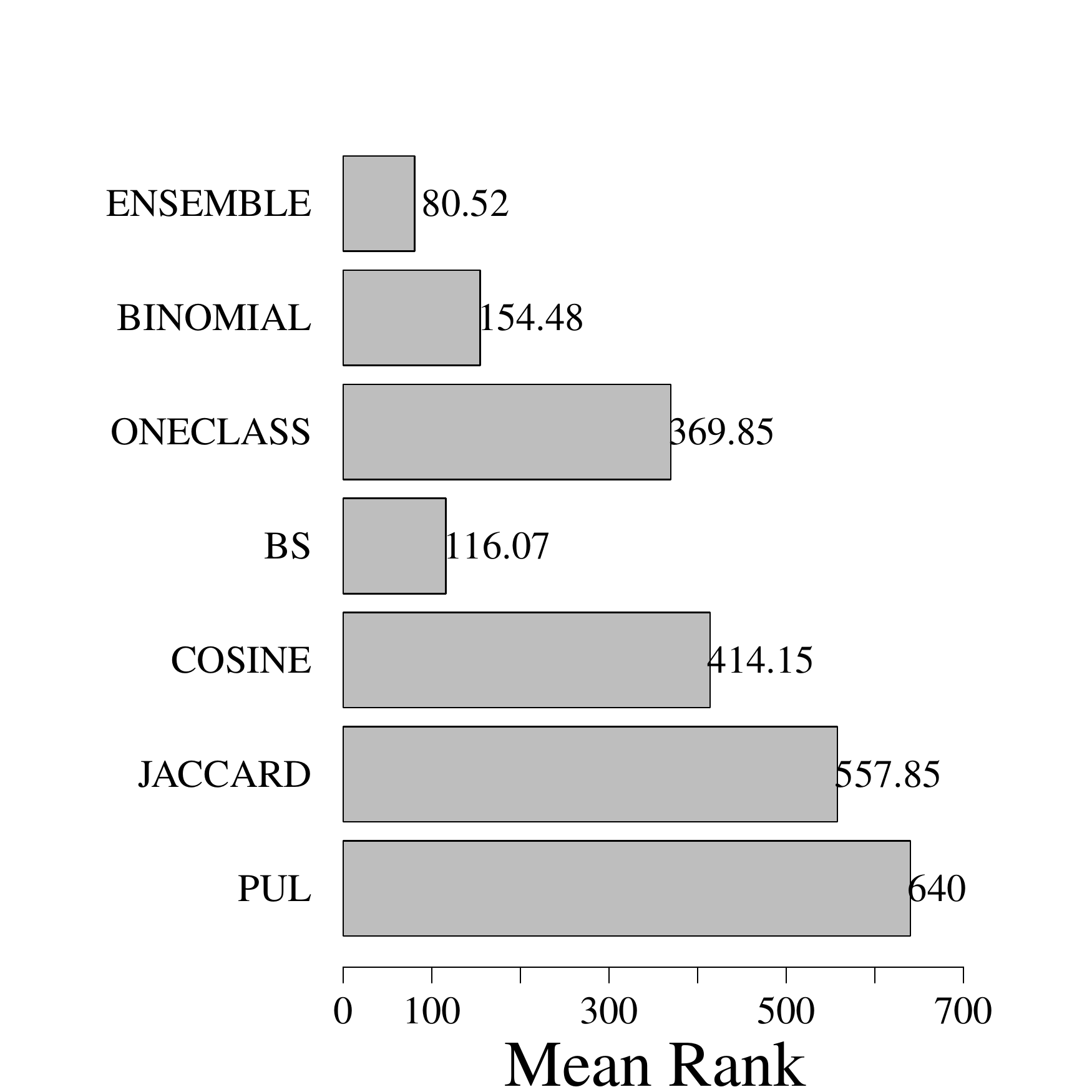}}
    \subfigure[SEC Domain]{
    \includegraphics[height=0.17\textheight]{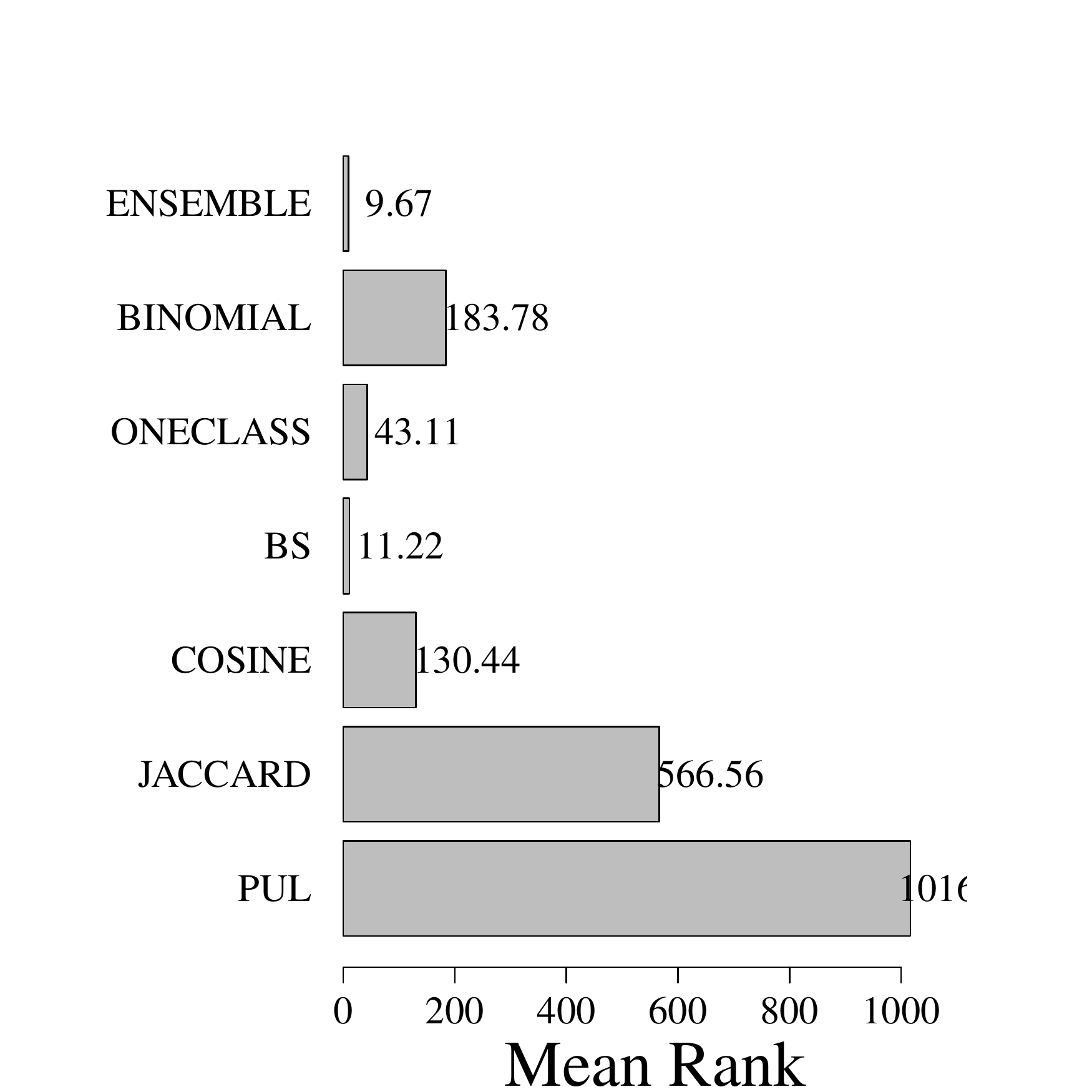}}
    \caption{Comparison of different ranking functions using mean rank metric in all domains. Lower values are better.}

    \label{fig:mean}
  \end{subfigure}

  \begin{subfigure}
    \centering
    \subfigure[Market Domain]{
    \includegraphics[height=0.17\textheight]{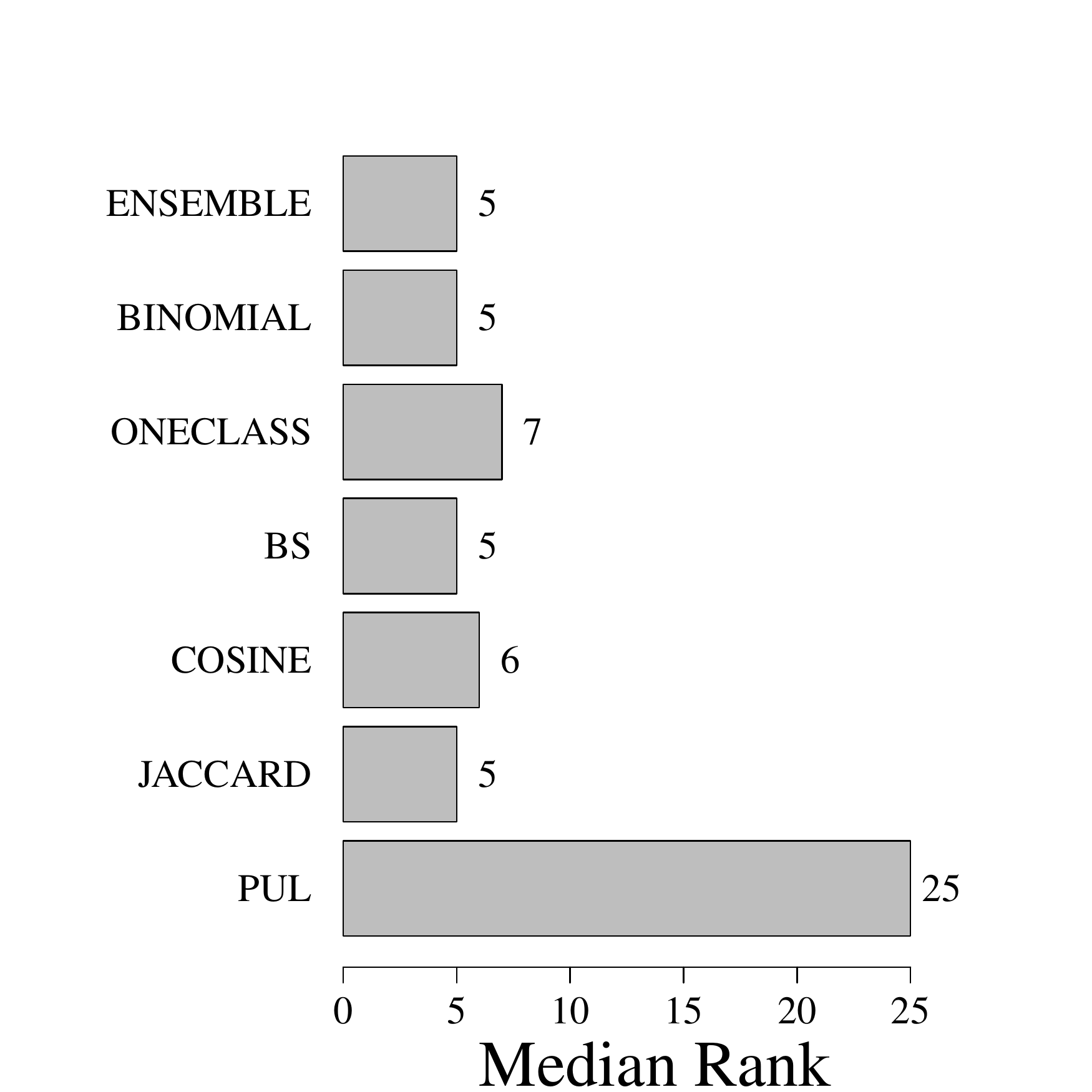}}
    \subfigure[Forum Domain]{
    \includegraphics[height=0.17\textheight]{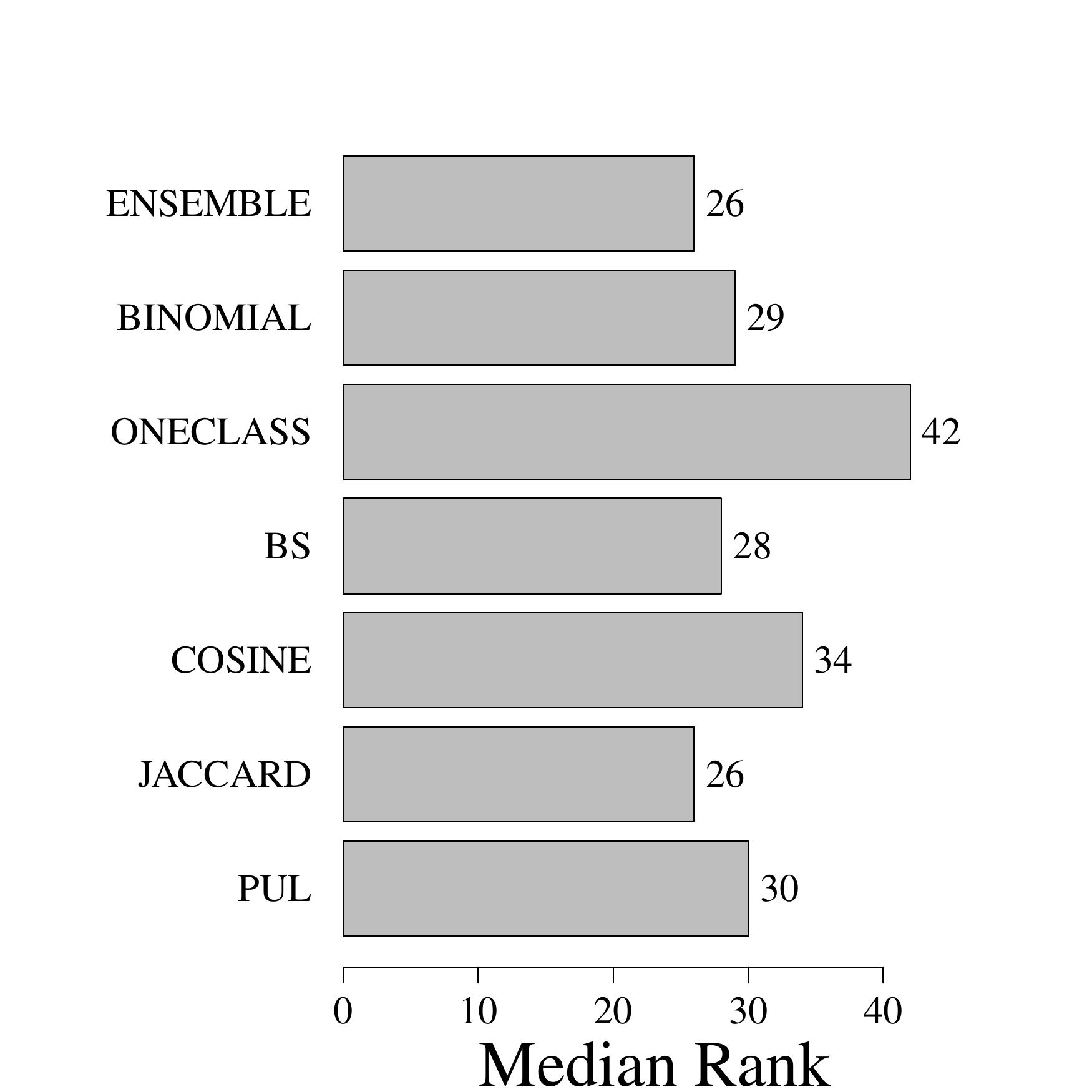}}
    \subfigure[HT Domain]{
    \includegraphics[height=0.17\textheight]{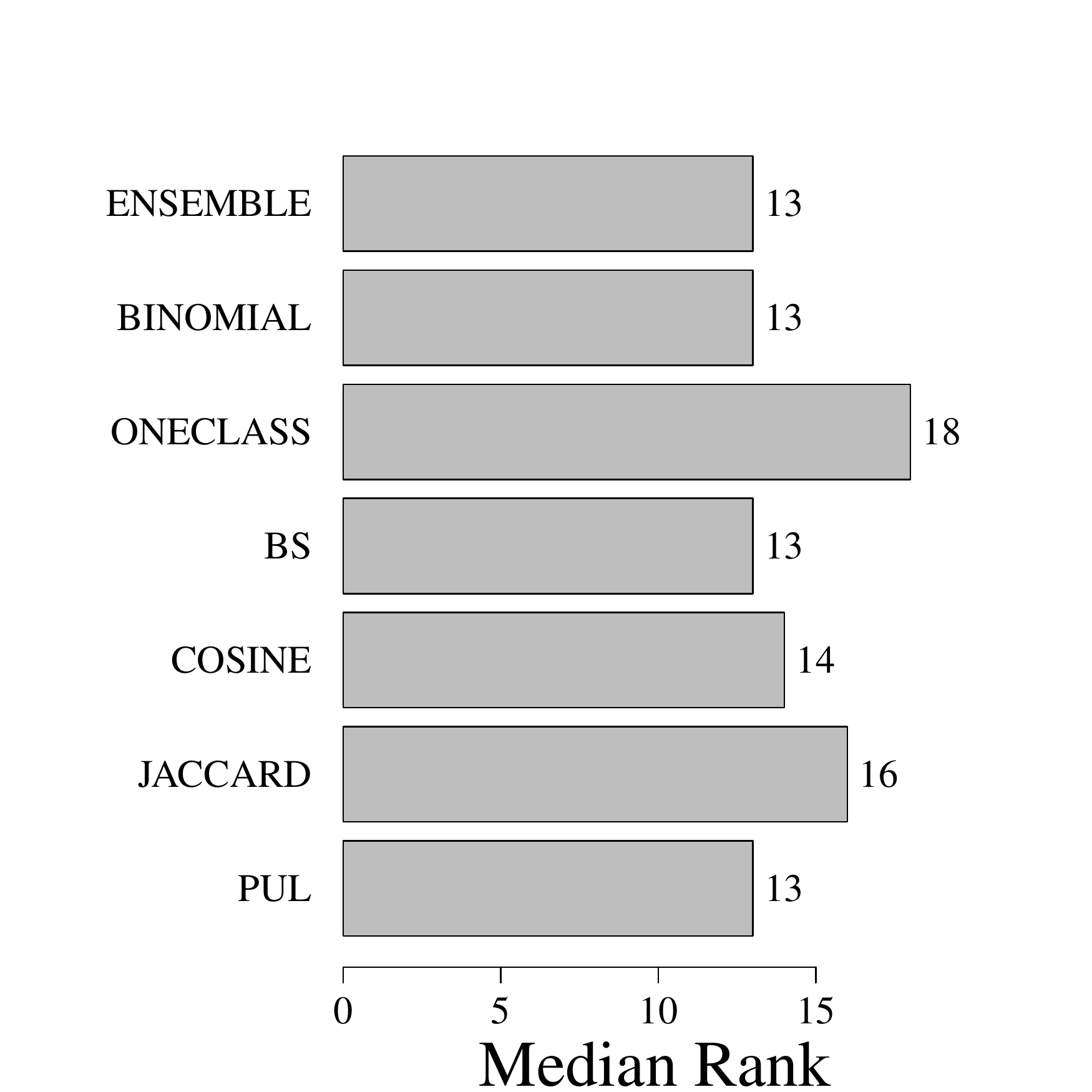}}
    \subfigure[SEC Domain]{
    \includegraphics[height=0.17\textheight]{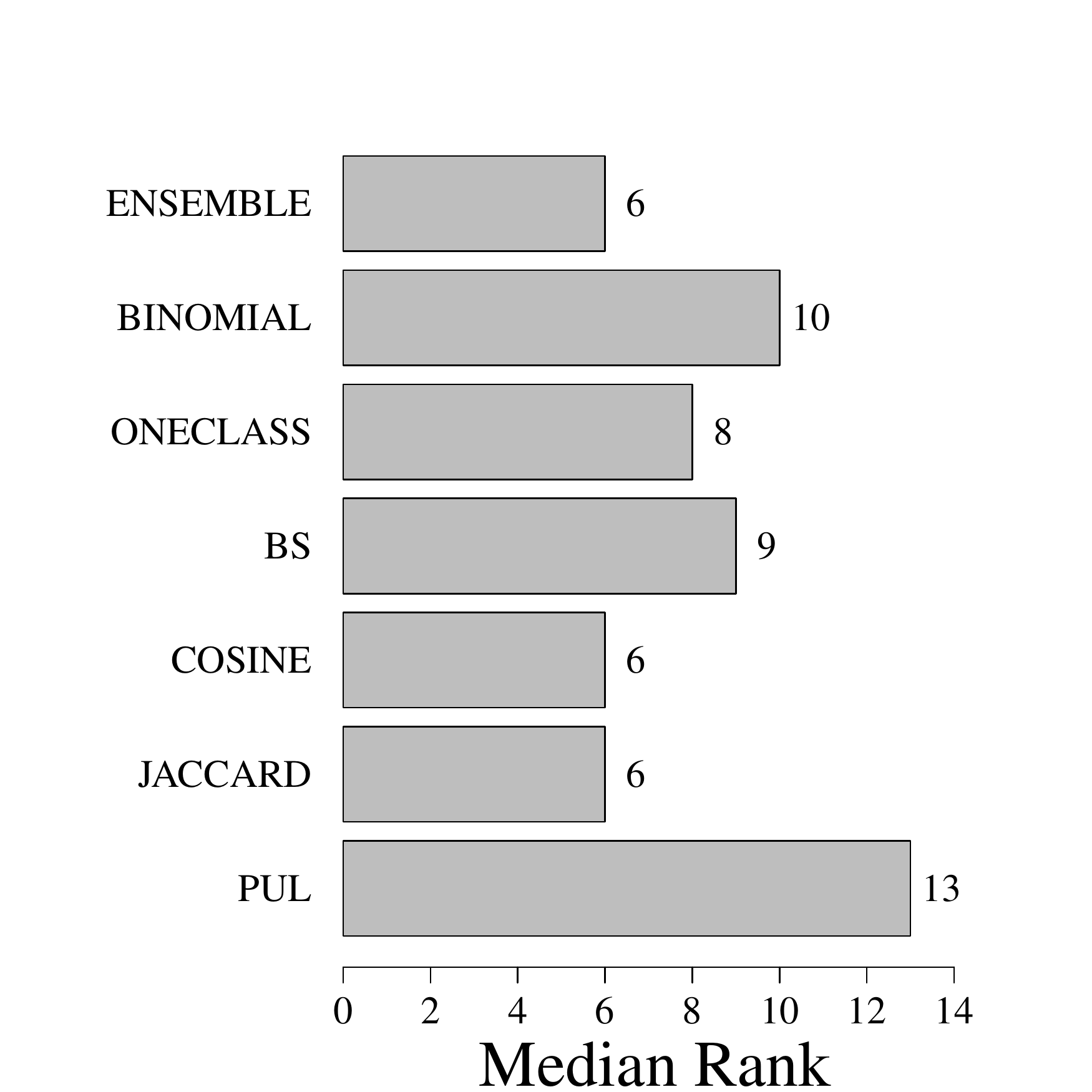}}
    \caption{Comparison of different ranking functions using median rank metric in all domains. Lower values are better.}

    \label{fig:median}
  \end{subfigure}

  \begin{subfigure}
    \centering
    \subfigure[Market Domain]{
    \includegraphics[height=0.17\textheight]{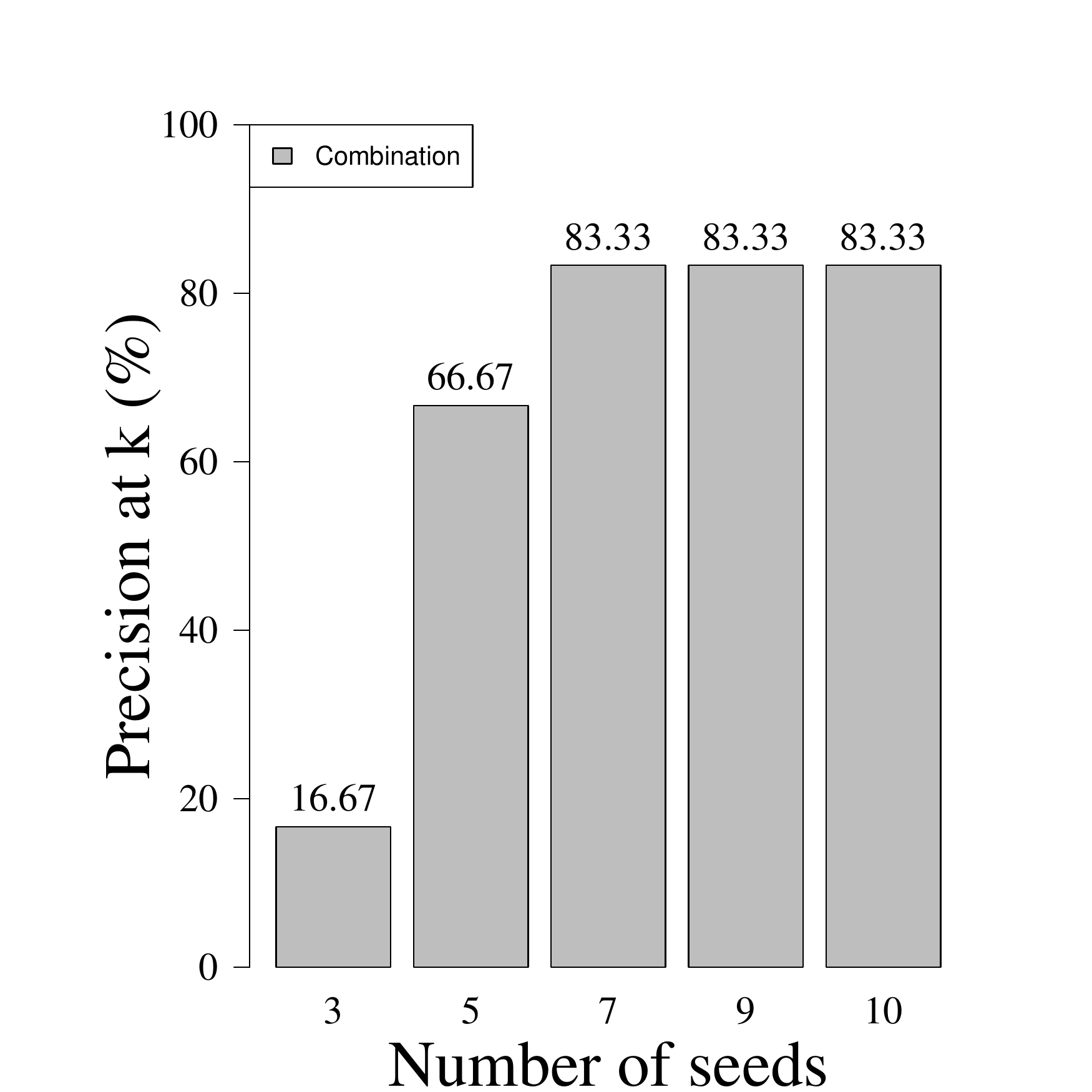}}
    \subfigure[Forum Domain]{
    \includegraphics[height=0.17\textheight]{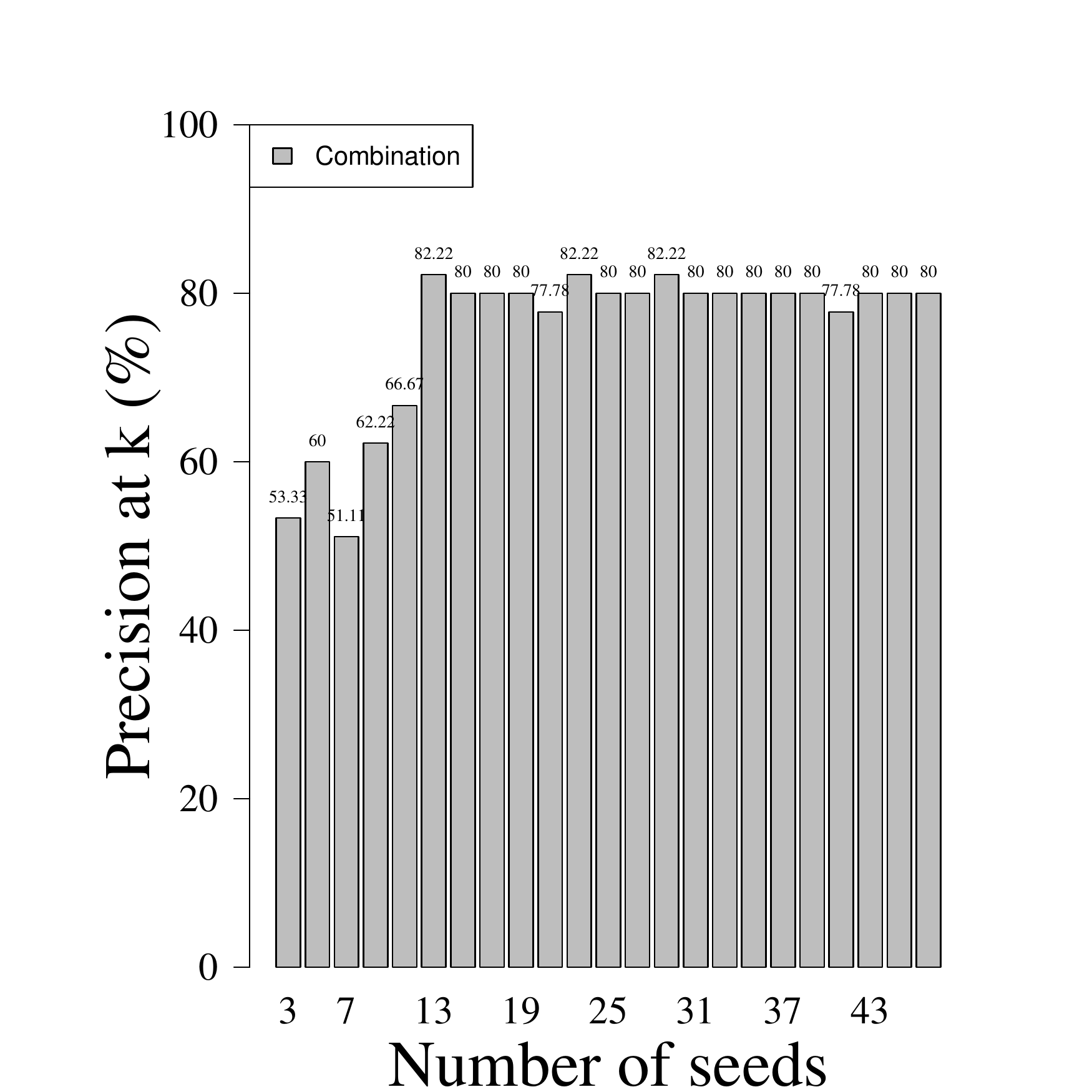}}
    \subfigure[HT Domain]{
    \includegraphics[height=0.17\textheight]{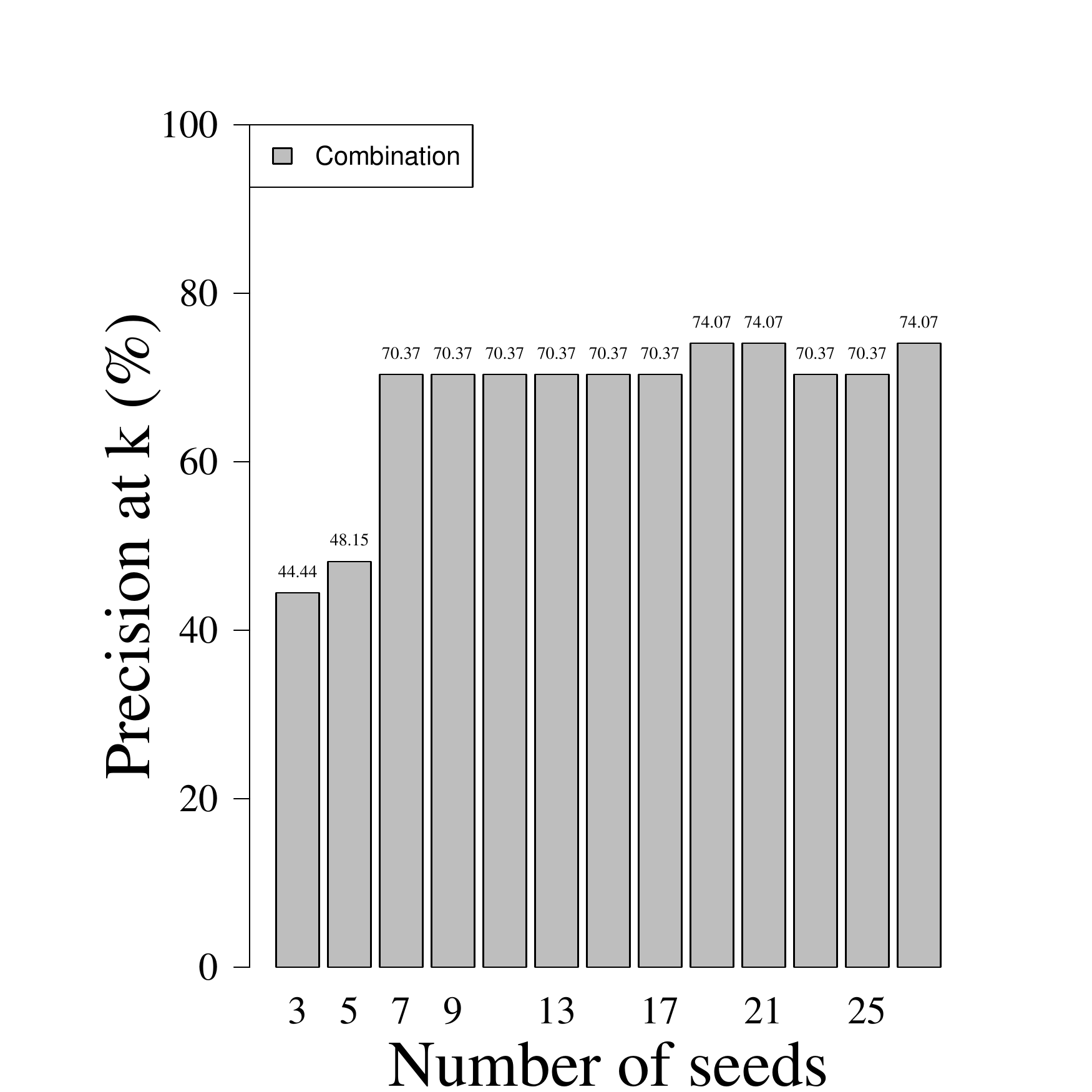}}
    \subfigure[SEC Domain]{
    \includegraphics[height=0.17\textheight]{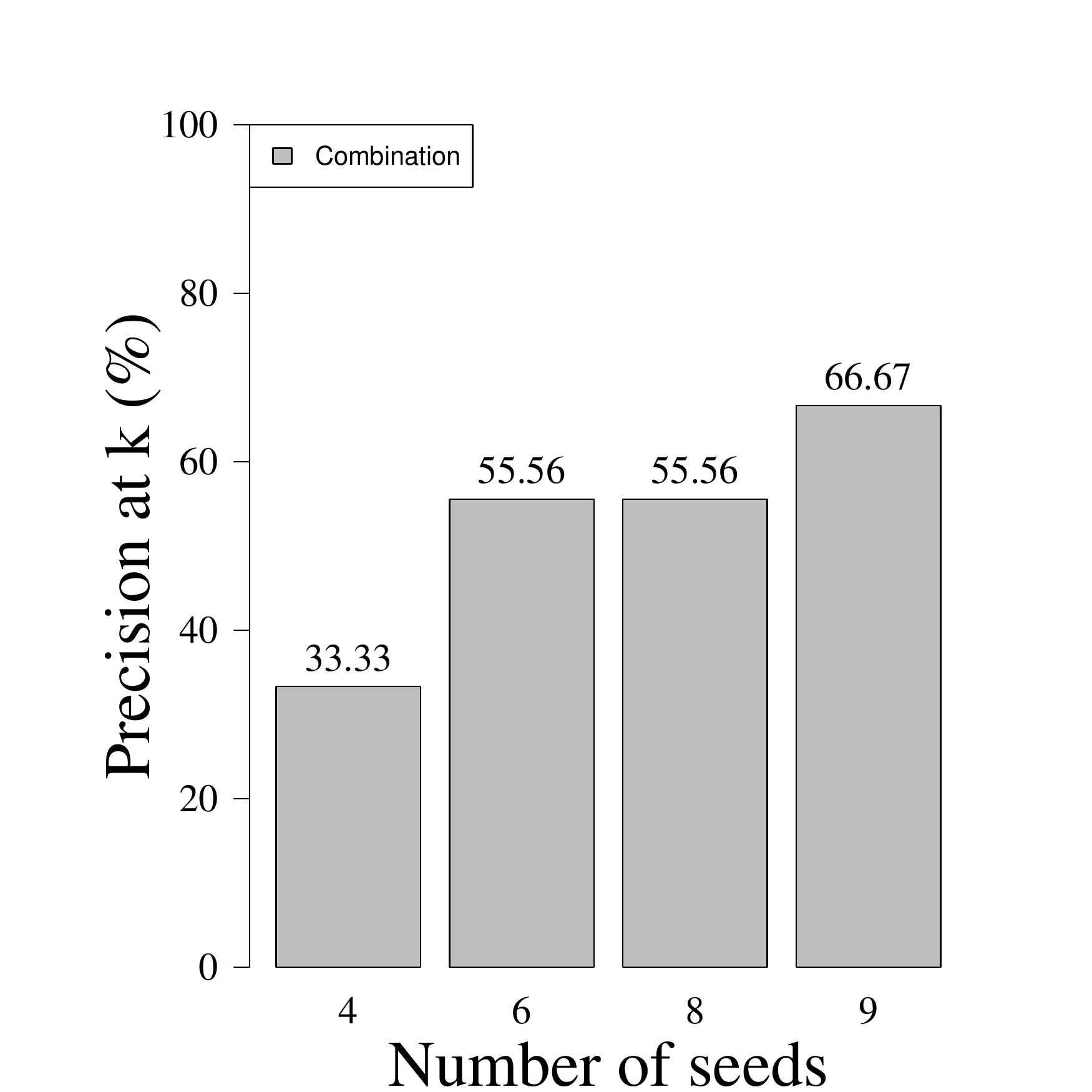}}
    \caption{Experiment showing the influence of the number of seed \websites. When the number of seeds increases, the precision also increases until it reaches a plateau. In general, less than 15 examples are required to attain high precision.}
  
    \label{fig:seeds}
  \end{subfigure}

\end{figure*}

Figures~\ref{fig:prec}, ~\ref{fig:mean}, and~\ref{fig:median}
compare the ranking functions using different metrics: $P@k$, mean rank and median rank.
Note that we select a $k$ that is equal to the number of relevant \websites (test set) in the candidate set.
By doing this, we basically normalize the $P@k$ values to range $[0,1]$.
The figure shows that the ENSEMBLE ranking function achieves better performance than the individual ranking functions do in most of the cases.
This confirms that the combination of different ranking functions is able to reduce the variance and increase the robustness of the overall ranking.
We also observe that \textit{BS} consistently performs better than
other individual ranking functions in median rank metric, i.e.,  it rarely places a relevant \website to the end of the list, which is one of our desired features.
Another observation is that the precision in SEC domain is relatively
low (66.67\%) compared to the others.  This can be explained by the
fact that in this domain, relevant \websites might contain a mix of
topics that are less distinguishable.  Another reason is that this
domain has smallest number of seeds among other domains.

\paragraph{Influence of the Number of Seed \Websites.}
To understand how the number of seed \websites impacts the effectiveness of the ranking,
we conduct a ranking evaluation with different number of seed \websites.
Figure~\ref{fig:seeds} shows the $P@k$ corresponding to number of seed \websites.
We notice that the precision improves when we increase the number of seed \websites up to 13 and stabilizes after that.
This finding suggests that this number of seeds is sufficient to achieve a good ranking performance.
It also reveals that probably there is still room to improve ranking performance,
especially for cases in which a larger number of seed \websites is available.
Given recent advances in text representation using dense vector representations
and document distance computation in the embedding space~\cite{kusner@ICML,wu@emnlp18},
this is a promising area for future work.
While this is a limitation of our ranking approach, it is still suitable for our context,
where we assume to have only a small list of seeds as input.

\vspace{-.3cm}
\subsection{Discovery Evaluation}
\label{sec:disc_eval}
In this section, we assess the effectiveness of DISCO and 
compare it to state-of-the-art discovery strategies~\cite{barbosa@www2007, barbosa@ijcnlp2011, vieira@wwwj2016} (baselines).
Since the baselines require a page classifier to operate, we train one
classifier for each domain using seed \websites and unlabeled examples
as training data.  Specifically, we consider seed \websites as
positive examples, and randomly select pages from DMOZ to use as
negative examples.  For training the classifiers, we use an
implementation of SVM from the scikit-learn
library~\cite{scikit-learn}.

To compute the discovery metrics, we train one ground-truth page
classifier for each domain using the positive and negative examples
manually labeled by SMEs.  Note that these classifiers are only used
for evaluation purposes.  Although not ideal, these classifiers are
the closest to the real ground truth that we can possibly obtain.
Nevertheless, they provide a scalable and consistent way to evaluate the discovery results.
Table~\ref{table:training_stat} summarizes the training data and cross-validation accuracy of these classifiers.
The accuracy ranges from 84\% to 97\%, which are sufficiently reliable for evaluating discovered \websites.
For each method, we stop the algorithm when it retrieves 50,000 \webpages.
In fact, all the baselines stop before reaching this limit.
In the following, we describe baselines and different configurations
of DISCO framework used in the  evaluation.

\begin{table}
  \centering
  \caption{Statistics of the training data and accuracy}
  \label{table:training_stat}
  \begin{tabular}{@{}cccc@{}}
  \toprule
  & \textbf{HT} &  \textbf{Market} & \textbf{Forum} \\
  \midrule
  \# of positive examples & 968 & 68 & 103 \\
  \# of negative examples & 7117 & 169 & 256 \\
  Accuracy (5-fold CV) & 0.97 & 0.96 & 0.84 \\
  \bottomrule
  \end{tabular}
\end{table}

We used the following approaches as baselines:

\paragraphbf{ACHE}:
ACHE is an adaptive focused crawler described in~\cite{barbosa@www2007}.
We used an open-source implementation available online.\footnote{http://github.com/ViDA-NYU/ache}
This implementation uses an online-learning crawling policy~\cite{soumen2002}
to prioritize the unvisited links.
Since ACHE is designed to maximize the number of relevant \webpages instead of relevant \websites,
we limit the number of crawled \webpages per \website to 10 to ensure it will
visit more \websites.
While a smaller limit would yield higher rate of new \websites,
it would potentially lead to missing relevant \websites that exist but would not have been discovered because some \webpages would never been visited.
Indeed, our experiment with a smaller limit results in a lower harvest rate
and the crawler also stops earlier due to running out of links to crawl.

\paragraphbf{BIPARTITE}:
This crawling strategy was proposed by Barbosa
et. al.~\cite{barbosa@ijcnlp2011} and, similar to ACHE, we set the page-per-site limit to 100.

\paragraphbf{SEEDFINDER}: We implemented the SeedFinder algorithm
proposed by Vieira et. al.~\citep{vieira@wwwj2016}.  SeedFinder uses the
keyword search operator to discover new \webpages. It then extracts new
keywords from the content of the pages classified as relevant by the
page classifier to derive new queries using relevance feedback.

We used the following search operators for DISCO:

\paragraphbf{KEYWORD}:
DISCO used with the keyword search operator.
For a fair comparison, this method uses the same settings as SEEDFINDER
(i.e., initial keywords and the number of search results returned by search APIs).

\paragraphbf{RELATED}:
DISCO used with the related search operator.
Since the objective is to discover \websites, we use host name of the input URL to form the search query.
For example, given the URL \texttt{http://example.com/path.html},
the query is represented as\\ \texttt{related:http://example.com}.

\paragraphbf{FORWARD}:
DISCO used with the forward crawling operator.

\paragraphbf{BACKWARD}:
DISCO used with the backward crawling operator.

\paragraphbf{BANDIT}:
DISCO used with the multi-armed bandits algorithm for selecting the discovery operator.

\begin{figure*}[!ht]
  \centering
  \subfigure[Market Domain]{
  \includegraphics[height=0.17\textheight]{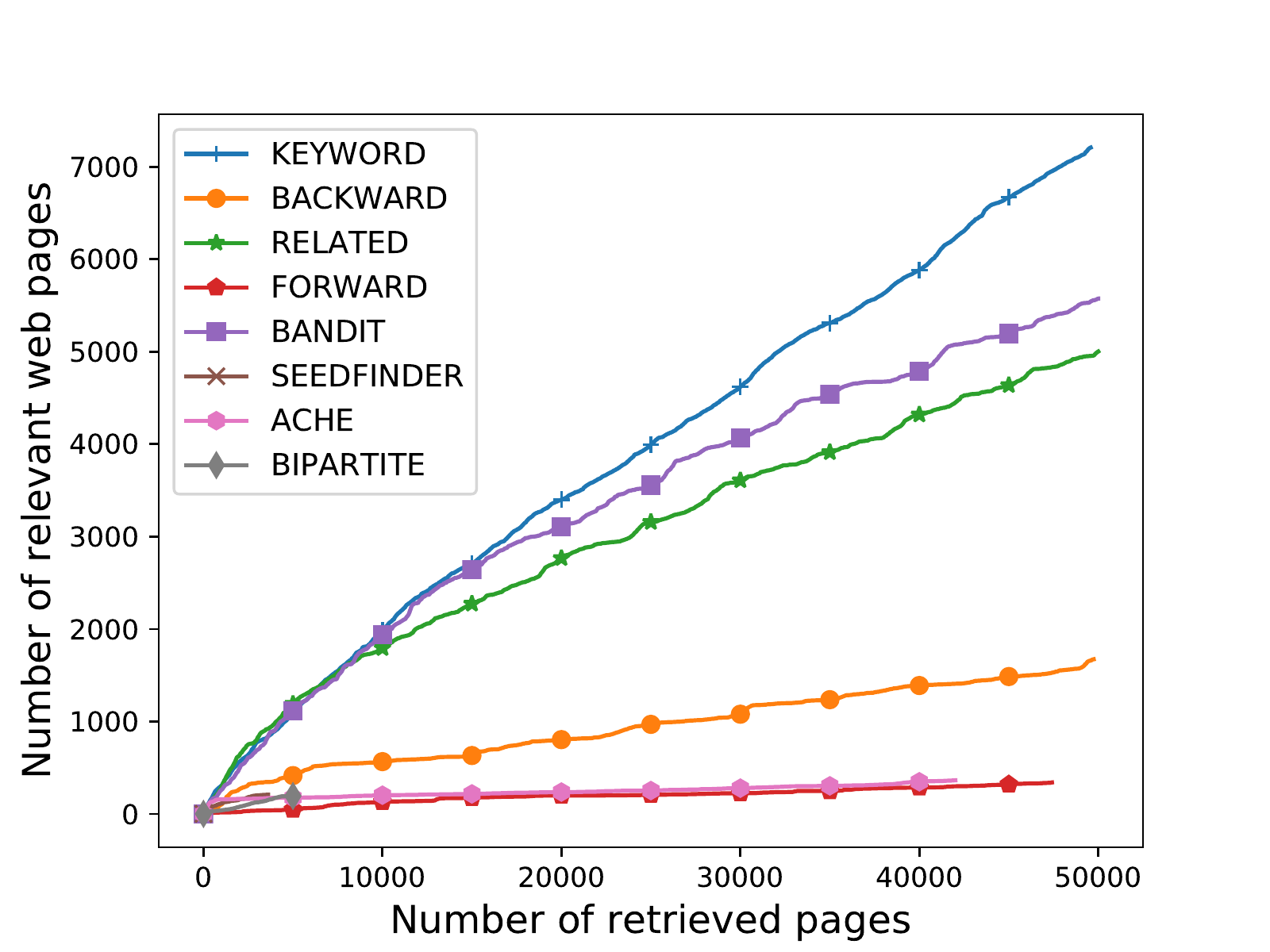}}
  \subfigure[Forum Domain]{
  \includegraphics[height=0.17\textheight]{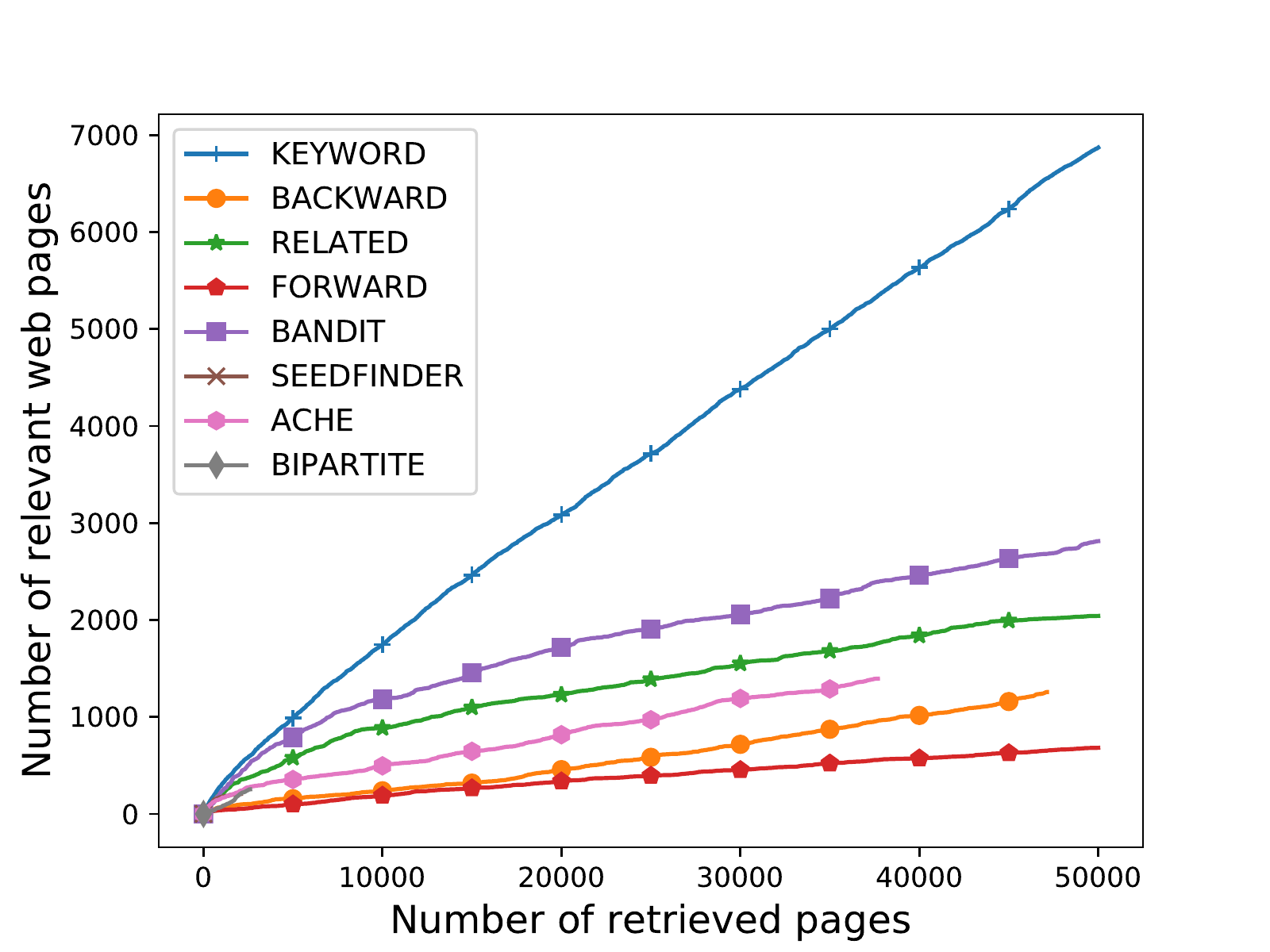}}
  \subfigure[HT Domain]{
  \includegraphics[height=0.17\textheight]{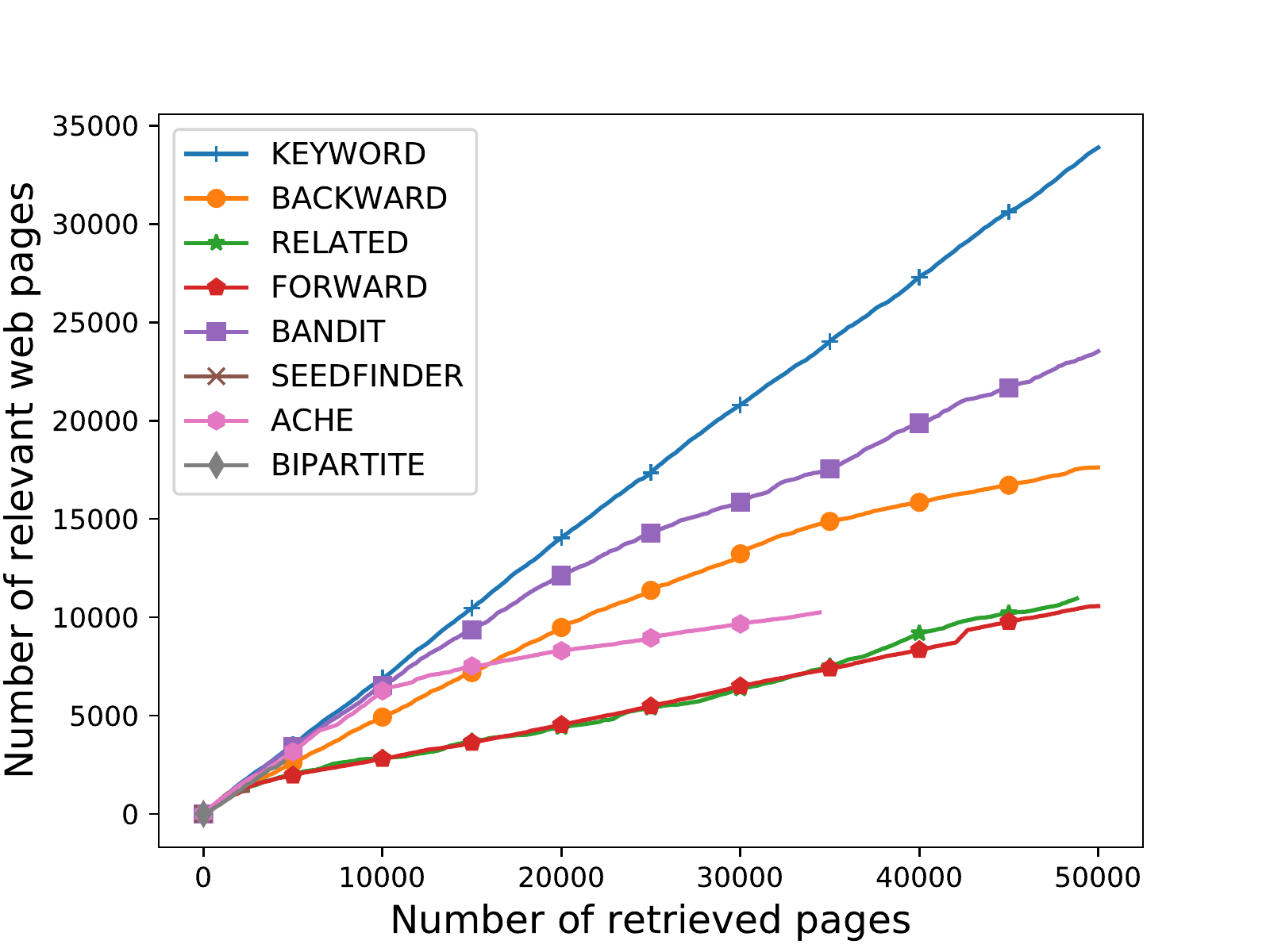}}
\vspace{-.15cm}
  \caption{Comparison of harvest rate of relevant \webpages for different discovery methods.}
\vspace{-.15cm}
  \label{fig:page_harvestrate}
\end{figure*}

\begin{figure*}[!ht]
  \centering
  \subfigure[Market Domain]{
  \includegraphics[height=0.17\textheight]{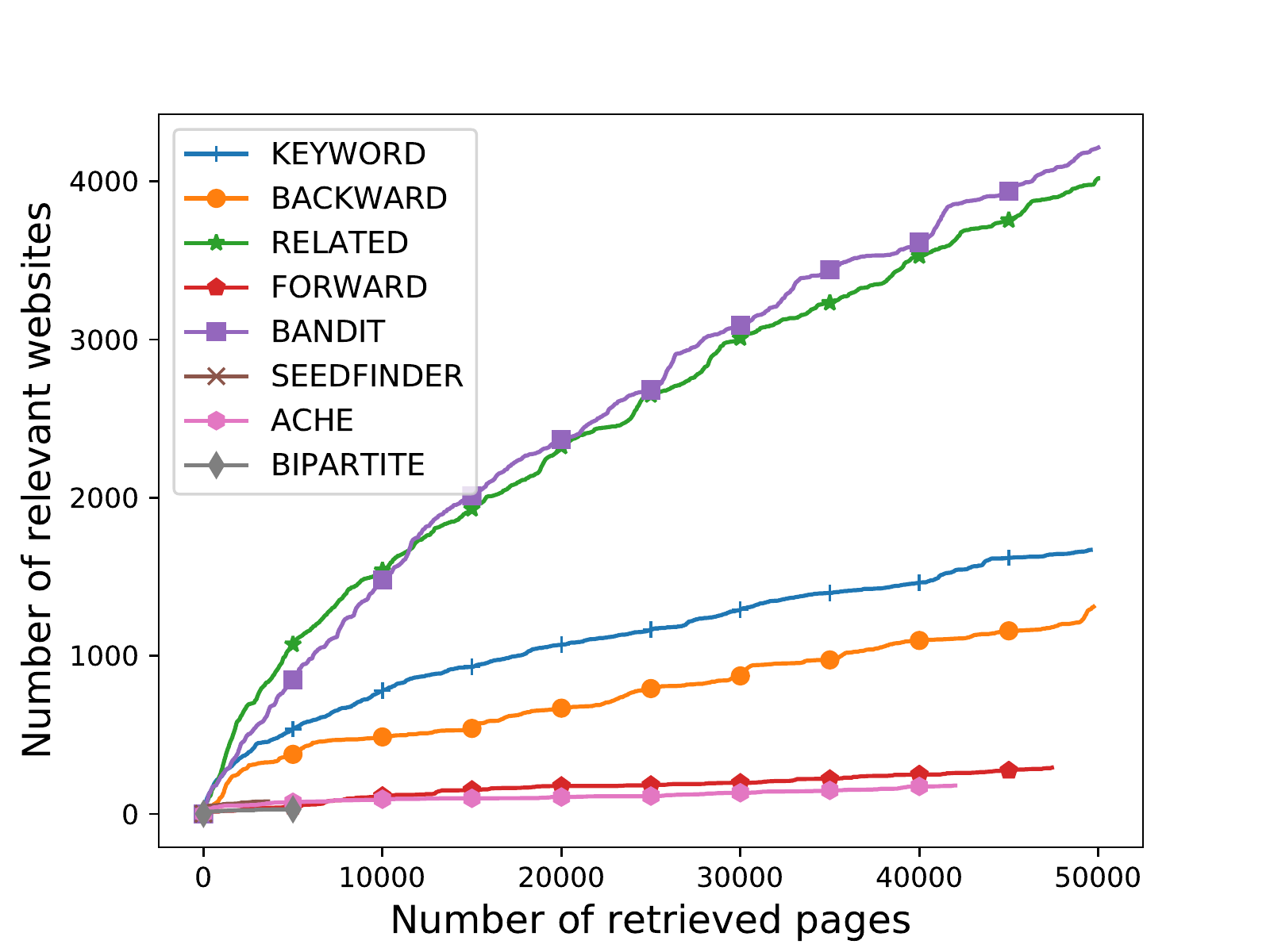}}
  \subfigure[Forum Domain]{
  \includegraphics[height=0.17\textheight]{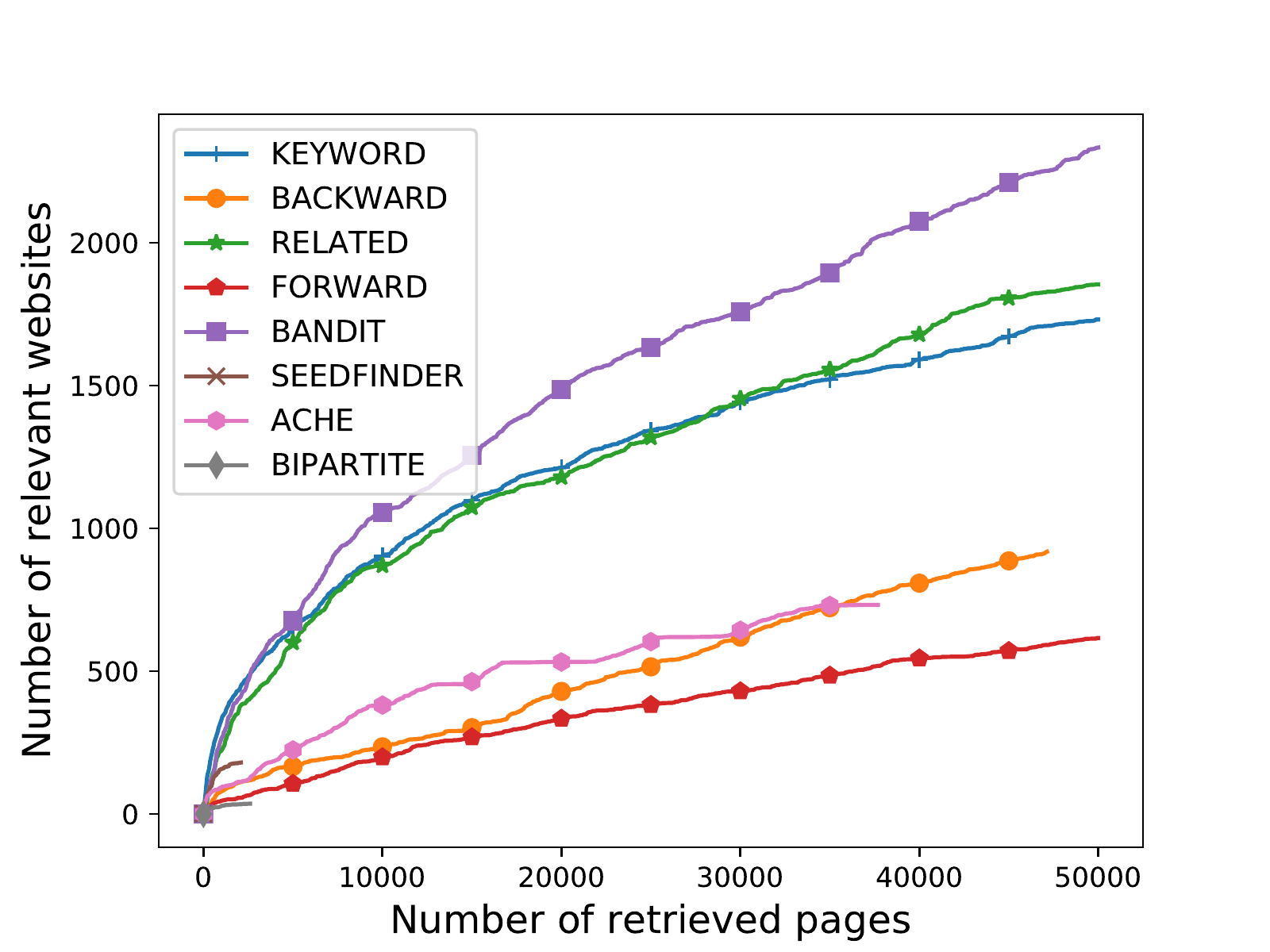}}
  \subfigure[HT Domain]{
  \includegraphics[height=0.17\textheight]{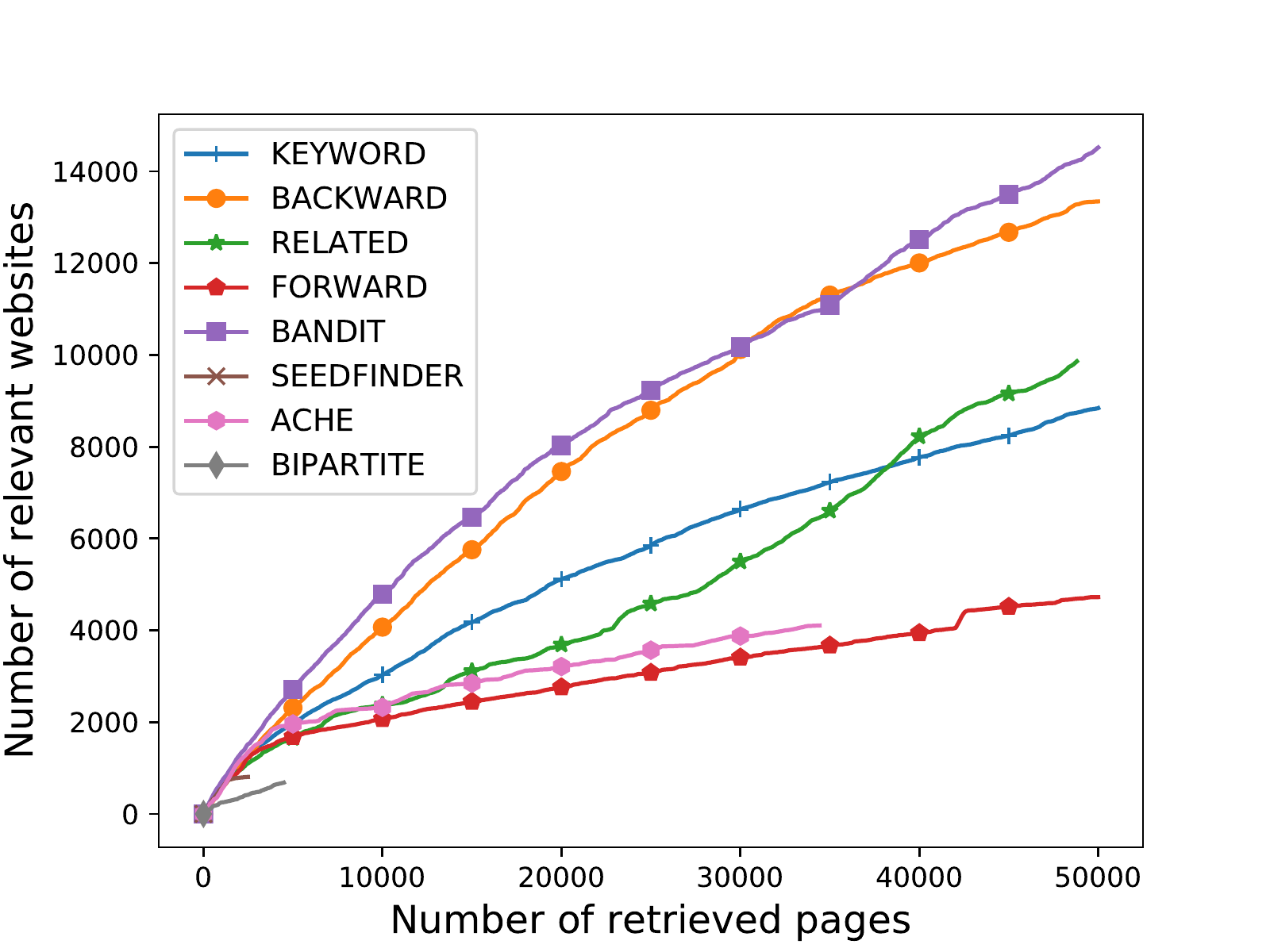}}
\vspace{-.15cm}
  \caption{Comparison of harvest rate of relevant \websites between for discovery methods.}
\vspace{-.15cm}
  \label{fig:site_harvestrate}
\end{figure*}

\begin{figure*}[!ht]
  \centering
  \subfigure[Market Domain]{
  \includegraphics[height=0.20\textheight]{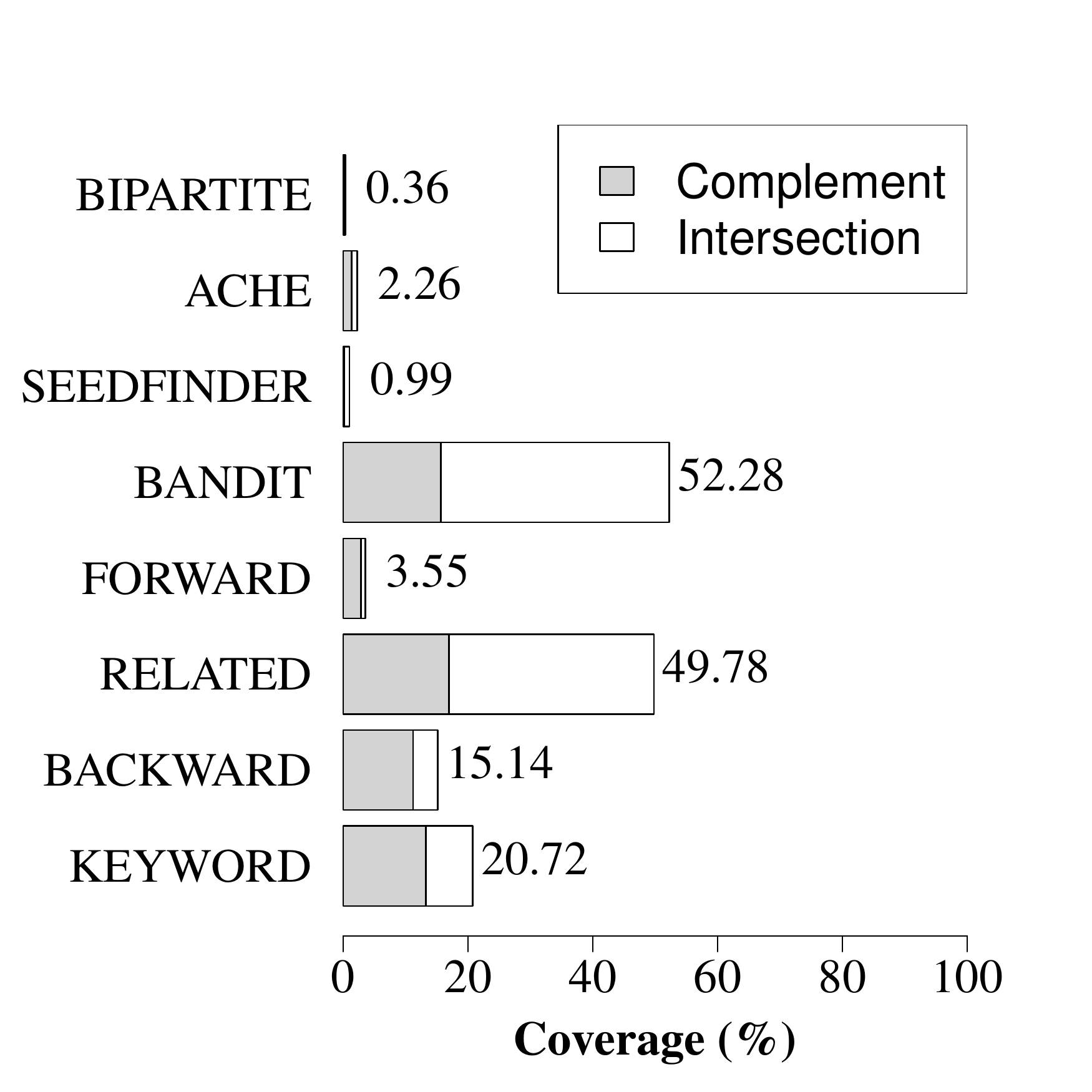}}
  \subfigure[Forum Domain]{
  \includegraphics[height=0.20\textheight]{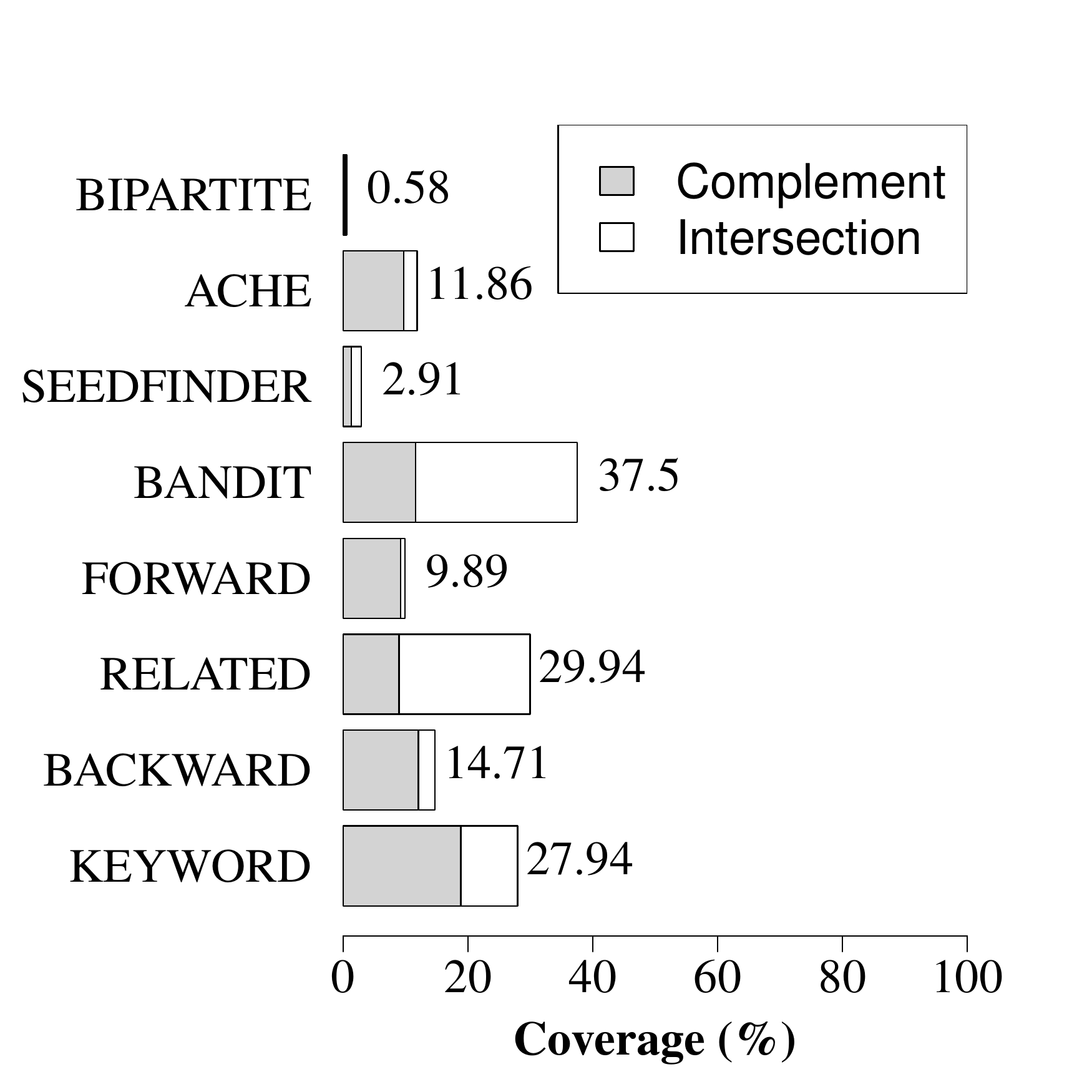}}
  \subfigure[HT Domain]{
  \includegraphics[height=0.20\textheight]{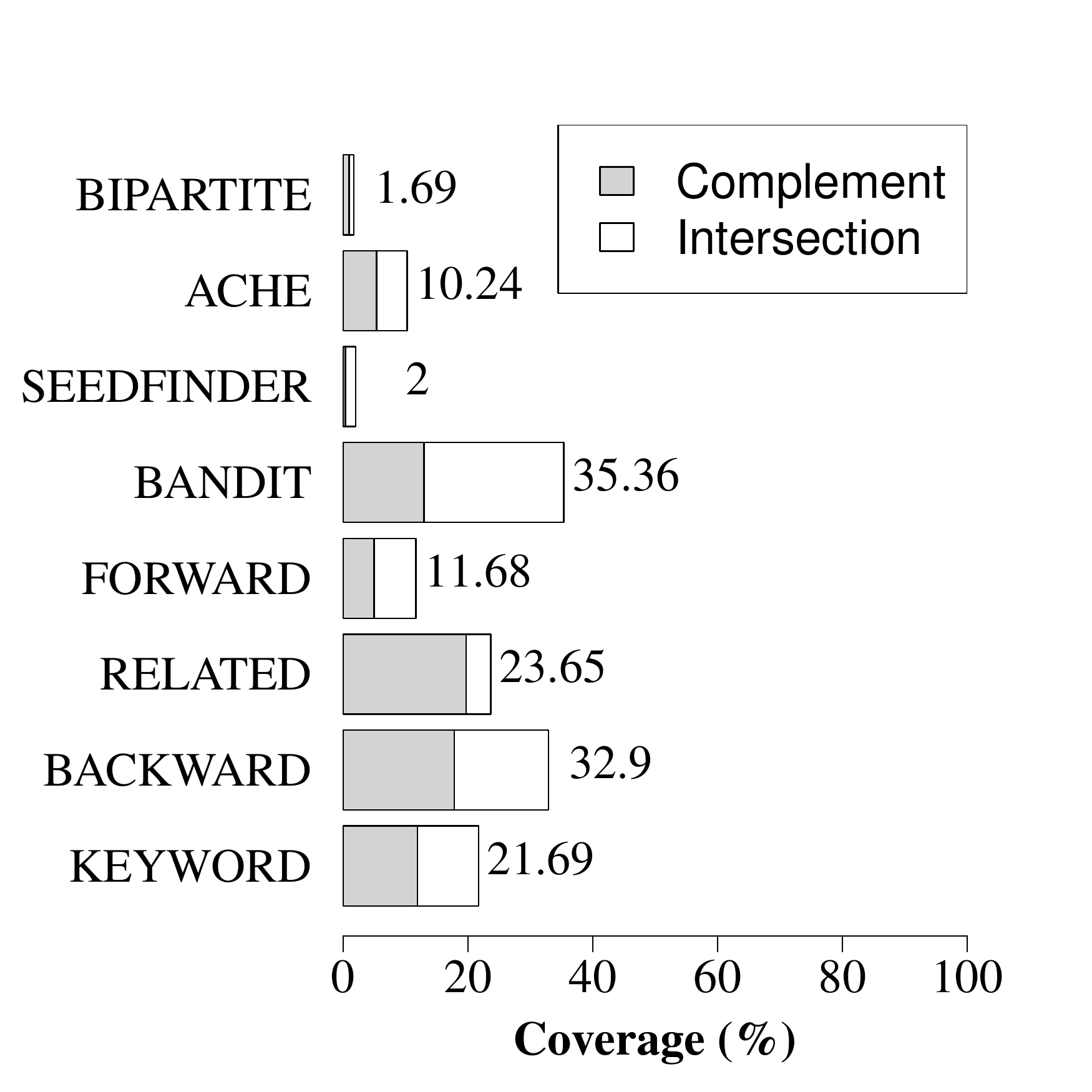}}
\vspace{-.15cm}
  \caption{Comparison of coverage for different discovery methods.}
\vspace{-.15cm}
  \label{fig:coverage}
\end{figure*}

\paragraph{Implementation Details.}
For  KEYWORD and SEEDFINDER methods, we use "gun classified", "gun forum" and "adult escort" as seed keywords
for the Market, Forum and HT domains respectively. We selected these
keywords since they represent well these domains and we observed
highly relevant results by searching these keywords on Google.
To obtain new keywords, we tokenize the textual content extracted from the description and keywords tag,
then select the most frequent keywords to use in the next search iteration. In each iteration,
we select at most 20 new keywords.
All methods start with the seeds provided by SMEs (also used in the ranking experiment).
We limit the number of results retrieved by keyword search, related
search and backlink search to be 50, 50, 5, respectively.
We set $k$ in Algorithm~\ref{alg:discovery} to 20.
DISCO uses ENSEMBLE as the ranking function because it outperforms other ranking functions
and is consistent across multiple domains.
We use MOZ API\footnote{https://moz.com/products/api} for backlink search,
Bing Web Search API\footnote{https://bit.ly/2KYAzu2} for keyword search,
and Google Custom Search API\footnote{https://developers.google.com/custom-search/} for related search.

\subsubsection{Comparing Harvest Rates}
We use the ground-truth classifiers to classify  the discovered \webpages.
A \website is considered as relevant if it contains at least one relevant \webpage.
Figure~\ref{fig:page_harvestrate} and Figure~\ref{fig:site_harvestrate} show
the number of discovered relevant \webpages and \websites versus the number of discovered \webpages over time.
Overall, DISCO with the best configuration (BANDIT) obtains at least
300\% higher harvest rate compared to other baselines.
Also, BANDIT consistently outperforms other configurations in all the three domains in terms of discovered \websites.
However, if we measure the harvest rate by relevant \webpages, ACHE performs closely to the DISCO strategies in the Market and Forum domains.
The reason is that ACHE's objective is to maximize the number of
relevant \webpages by exploiting relevant \websites that are already
discovered.
However, it quickly reaches a plateau when fewer relevant \websites are found.
Another reason that ACHE performs well in HT domain is that \websites in this domain are highly connected.
This also explains why backlink method performs better in this domain
than the others.  SEEDFINDER and BIPARTITE strategies stop early,
before reaching 5,000 pages in all the domains, due to running out of
links to crawl.

\subsubsection{Comparing Coverage}
Since it is impractical to obtain the real ground truth for each domain (i.e., all the relevant \websites in the domain),
we use the union of discovered \websites from all discovery methods as
an approximation for the ground truth.
By doing this, while the estimated coverage can be largely different from the real value,
it serves as reasonable relative measure to compare different methods.
Figure~\ref{fig:coverage} shows the coverage of all discovery methods in the considered domains.
For each method, we compute the percentage of relevant \websites that are also discovered by all other methods,
which we denominate as \textit{intersection} in the plots.
The \textit{complement} region represents the percentage of relevant
\websites that are discovered only by the corresponding method.
The BANDIT strategy consistently attains the largest intersection among all methods.
This suggests that each operator searches a different region of the Web,
and by adaptively combining them, the BANDIT strategy is able to explore a broader portion of the Web.

\section{Conclusion}
\label{sec:conclusion}
In this paper, we propose DISCO, a new approach that helps bootstrap
domain discovery: given a small number of sample sites in a domain, it
automatically expands the set. 
DISCO employs an iterative crawling procedure that combines multiple
search operators and, as proxy to select relevant pages, it ranks the
search results based on their similarity to the sample sites.
We perform extensive experiments using real-world data from different social-good
domains and show that our framework is effective and attains significant gains in harvest rate
compared with state-of-the-art methods. The experimental results
suggest that DISCO can be effective at lowering the burden on users that need to
create page classifiers and discover information on the Web at
scale. For example, DISCO could be used to both automatically provide additional seed
pages for DDT~\cite{krishnamurthy@kdd2016}  and to rank the results of
user-specified search operations.

While DISCO is effective for the domains we experimented with, it has
limitations. 
Our ranking approach is only suitable for the domains for which using n-gram features are effective
and the given seed websites are representative for the  domains.
Additionally, keyword search is very sensitive to the seed keywords, which can be hard to select 
for some domains.
In future work, we would like to explore image-based search for this
problem, which can be potentially applied to image-rich domains, such
as weapon ads and human trafficking.

\vspace{10pt}
\myparagraph{Acknowledgments.}
This work was partially supported by the DARPA MEMEX and D3M programs,
and NSF award CNS-1229185.
Any opinions, findings, and conclusions or recommendations expressed in
this material are those of the authors and do not necessarily reflect
the views of NSF and DARPA.

\balance

\bibliographystyle{ACM-Reference-Format}
\bibliography{paper}

\end{document}